\documentclass[11pt,a4paper]{article}
\usepackage{graphicx}
\usepackage{jcappub}
\usepackage{subfig}
\usepackage[latin1]{inputenc}
\usepackage{amsmath}
\usepackage{amsfonts}
\usepackage{amssymb}
\usepackage{setspace}
\usepackage{mathtools}
\usepackage{pstricks}
\usepackage{color}
\providecommand{\f}[2]{\frac{{#1}}{{#2}}}
\newcommand{\da}{\ensuremath{\dot{a}}}
\newcommand{\dda}{\ensuremath{\ddot{a}}}

\newcommand{\ee}[1]{\begin{equation}#1\end{equation}}
\newcommand{\ea}[1]{\begin{align}#1\end{align}}
\newcommand{\eg}[1]{\begin{gather}#1\end{gather}}
\newcommand{\g}{\gamma}

\title{A Simple Method for One-Loop Renormalization in Curved Space-Time}

\author[a]{Tommi Markkanen,}
\author[b,c]{Anders Tranberg}

\affiliation[a]{Helsinki Institute of Physics and Department of Physics, P. O. Box 64, FI-00014, University of Helsinki, Finland.}
\affiliation[b]{Niels Bohr International Academy and Discovery Center, Niels Bohr Institute,\\  Blegdamsvej 17, 2100 Copenhagen, Denmark}
\affiliation[c]{Faculty of Science and Technology, University of Stavanger,
N-4036 Stavanger, Norway} 

\emailAdd{tommi.markkanen@helsinki.fi}
\emailAdd{anders.tranberg@uis.no}

\abstract{We present a simple method for deriving the renormalization counterterms from the components of the energy-momentum tensor in curved space-time. This method allows control over the finite parts of the counterterms and provides explicit expressions for each term separately. As an example, the method is used for the self-interacting scalar field in a Friedmann-Robertson-Walker metric in the adiabatic approximation, where we calculate the renormalized equation of motion for the field and the renormalized components of the energy-momentum tensor to fourth adiabatic order while including interactions to one-loop order. Within this formalism the trace anomaly, including contributions from interactions, is shown to have a simple derivation. We compare our results to those obtained by two standard methods, finding agreement with the Schwinger-DeWitt expansion but disagreement with adiabatic subtractions for interacting theories.}


\begin{document}
\maketitle

\section{Introduction}
\label{sec:introduction}
Quantum field theory in curved space-time has proven to be successful for studying the cosmological implications of quantum fields, especially when interactions can be neglected \cite{ParkerToms, birreldavies}. Other areas of physics have also benefited from the use of quantum fields in curved space \cite{Unruh:1994je}. Only recently have the effects of interactions made their way into such considerations. It is well-known that renormalization of interacting theories gives rise to important phenomena, such as the running of the couplings, and in order to derive the quantum corrected equations of motion, one must be able to perform consistent renormalization in curved space-time. The early work in this field concentrated on the consistent cancellation of divergences, without explicitly calculating the finite remainder of the counterterms \cite{Birrell:1978zs,Bunch:1979uk, Bunch:1980br,Bunch:1980bs,Birrell:1979ir,Toms:1982af}. However, if one is interested in quantitative physical predictions, the finite parts of the counterterms, in particular the renormalization scale, must be known in order to assign a proper physical interpretation for the constants of the theory.

The various approaches to this problem can roughly be divided into two categories: Renormalizing at the level of the action and renormalizing at the level of the equations of motion. In the latter approach, the most popular method has been adiabatic subtraction \cite{Parker:1974qw,Fulling:1974zr, Fulling:1974pu} (see also \cite{MolinaParis:2000zz}), having recently been applied in \cite{Ramsey:1997sa, Anderson:2005hi, tranberg}. The former approach has also been found fruitful \cite{Hu:1984js,Kirsten:1993jn, Elizalde:1994ds,Buchbinder:1984ww,Buchbinder:1985js,Buchbinder:1986yh,shapiro2003,Markkanen:2012rh}. 

The ground work for renormalizing the quantum corrected action was laid in the seminal paper \cite{Coleman:1973jx} and can be generalized to curved spaces \cite{Toms:1982af}. Since the energy-momentum tensor is calculated as a variation with respect to $g^{\mu\nu}$, the effective action must be derived with a general metric. This is a highly involved task and there is currently quite a limited number of methods for deriving the effective action in curved space. Amongst the most popular is the gradient expansion, also known as the Schwinger-DeWitt expansion \cite{Minakshisundaram:1949xg,DeWitt:1965jb}, which was generalized in \cite{Barvinsky:1985an}. The gradient expansion works well for calculating the divergent contribution to the counterterms, but it is only reliable for the complete effective action when the fields are slowly varying, thus having limited applicability in many physical scenarios. Another proposed method can be found in \cite{Barvinsky:1993en} (and references therein) but it is mathematically involved.

The major benefit of adiabatic subtraction is that because working at the equation of motion level, one is free to constrain the metric, which in many scenarios offers a significant simplification over working with a general space-time. However, adiabatic subtraction does suffer from a disadvantage: If one does not introduce a procedure for also fixing the finite parts of the counterterms, the renormalization conditions for all the coupling constants is left implicit leaving the physical interpretation of these constants ambiguous. It has been shown that for the non-interacting scalar theory, adiabatic subtraction is equivalent to redefining the constants of the original action \cite{Bunch:1980vc}. We are not aware of work examining the validity of adiabatic subtraction for interacting theories; whether the renormalization procedure can be reduced to the introduction of counterterms in the usual way.

The aim of this work is to present a consistent and systematic renormalization procedure working at the level of the equations of motion, with explicit renormalization conditions for each constant and control over the various finite parts of the counterterms. This method combines the benefits of adiabatic subtraction and the effective action approach. We will apply this method in the context of the self-interacting scalar field, i.e. a $\lambda\varphi^4$-theory, in a Friedmann-Robertson-Walker (FRW) space-time with zero spatial curvature, calculated to one-loop order. Explicit results will be derived for the energy-momentum tensor and the equation of motion for the $\varphi$-field in the fourth order adiabatic vacuum. The regularization method will be dimensional regularization. The conformal anomaly, including contributions from interactions, is shown to have a simple derivation within this formalism. Finally, the results are compared to those obtained by the gradient expansion and adiabatic subtraction.

The paper is organized as follows. In section \ref{sec:obtaining} we discuss renormalization in curved space-time, with special attention being given to renormalization of the effective action and adiabatic subtraction. Section \ref{sec:Field equations} contains the derivations for the one-loop equations of motion for the field and the metric in the adiabatic vacuum. In section \ref{sec:expli} we present the explicit results for the equations of motion and the conformal anomaly. Comparison between our results and those derived via the gradient expansion and adiabatic subtraction is done in section \ref{sec:comparison} and we conclude in section \ref{sec:Summary}. We will be using natural units where $c=\hbar=1$ and the $(+,+,+)$ convention of \cite{Misner:1974qy} for the various tensors. Bare quantities are denoted with a subscript "$0$", $n-1$ dimensional spatial parts of vectors with a boldface letter and operators with a "hat", $\hat{~~}$.

\section{Obtaining renormalized equations of motion}
\label{sec:obtaining}

Quantum field theory in curved space-time is a theory of quantized fields on a curved classical background \cite{ParkerToms, birreldavies}. In the functional integral approach this means that the generating functional $Z[J]$ has path integration over only the matter fields. For a theory with a single field $\varphi$ and a classical action $S[\varphi,g^{\mu\nu}]$ this gives
\ee{Z[J]=\int \mathcal{D}\varphi~e^{iS[\varphi,g^{\mu\nu}]+i\int d^4x\sqrt{-g}~J\varphi}.}

The quantum corrected or effective action denoted as $\Gamma[\varphi,g^{\mu\nu}]$, can be derived by a Legendre transformation \cite{{Peskin:1995ev}}
\ee{\Gamma[\varphi,g^{\mu\nu}]\equiv\int d^4x \sqrt{-g}~\mathcal{L}_{eff}[\varphi,g^{\mu\nu}]\equiv-i\log Z[J]-\int d^4x\sqrt{-g}~J\varphi,}
where it must be borne in mind that now $\varphi$ represents the expectation value of the field, $\langle\hat{\varphi}\rangle\equiv \varphi$. In general the effective action is divergent and renormalization must be implemented in order to have a meaningful result. For a renormalizable theory, the removal of the divergences is achieved by appropriately tuning the constants introduced by the original action $S[\varphi,g^{\mu\nu}]$. After renormalization is done for $\Gamma[\varphi,g^{\mu\nu}]$, the finite equations of motion for the field and the metric can be derived via simple variation as
\ee{\label{eq:var}\f{\delta\Gamma[\varphi,g^{\mu\nu}]}{\delta\varphi(x)}=0,\quad\f{\delta\Gamma[\varphi,g^{\mu\nu}]}{\delta g^{\mu\nu}(x)}=0.}

One could equally well use the quantized action $S[\hat{\varphi},g^{\mu\nu}]$ to derive the operator equations of motion
\ee{\label{eq:var2}\bigg\langle\f{\delta S[\hat{\varphi},g^{\mu\nu}]}{\delta\hat{\varphi}(x)}\bigg\rangle=0,\quad\bigg\langle\f{\delta S[\hat{\varphi},g^{\mu\nu}]}{\delta g^{\mu\nu}(x)}\bigg\rangle=0}
and renormalize the equations for the expectation values after the variation. Again for a renormalizable theory, we only need to adjust the constants of the action in order to render these equations finite. The first equation in (\ref{eq:var2}) is the Schwinger-Dyson equation for the one-point function\footnote{In this language, we will be truncating at one-loop, using bare vertices and propagators.} \cite{Peskin:1995ev}. 

In this paper our classical action will be\footnote{At this level, all bare gravitational coefficients are written with a positive sign. They can later be matched to normal conventions, where for instance $\Lambda\rightarrow -\Lambda/( 8 \pi G) $, $\alpha =1/(16\pi G)$ and so forth.}
\ea{S[\varphi,g^{\mu\nu}]&\equiv S_m[\varphi,g^{\mu\nu}]+S_g[g^{\mu\nu}]\nonumber \\\label{eq:actm}S_m[\varphi,g^{\mu\nu}]&=-\f{1}{2}\int d ^4x\sqrt{-g}~ \bigg[\partial_\mu\varphi\partial^\mu\varphi+m^2\varphi^2 +\xi R\varphi^2+2\f{\lambda}{4!}\varphi^4\bigg]\\\label{eq:actg}S_g[g^{\mu\nu}]&= \int d^4x\sqrt{-g}~\bigg[\Lambda+\alpha R+\beta R^2+\epsilon_{1}R_{\alpha\beta}R^{\alpha\beta}+ \epsilon_{2}R_{\alpha\beta\gamma\delta}R^{\alpha\beta\gamma\delta}\bigg]
.}
We have included $\mathcal{O}(R^2)$ tensors, which although not present in the classical Einstein-Hilbert action, are required in the quantized theory for consistently cancelling all the divergences in the energy-momentum tensor \cite{birreldavies}.

\subsection{Renormalization via the effective action}
\label{sec:effective}

Given a regularized expression for the effective action $\Gamma[\varphi,g^{\mu\nu}]$, renormalization can be acquired with relative ease: Each counterterm is determined by matching the effective action to the classical one, for each constant separately at some scale choices for the field, its derivative and the gravitational quantities. For an example, see \cite{Markkanen:2012rh}. In general this procedure must be used for all the constants of the classical theory and one must supply a scale for each degree of freedom. For the theory defined by (\ref{eq:actm} - \ref{eq:actg}) with a mean field that has a dependence only on time, the renormalization conditions are 
\ea{\f{\partial^2\mathcal{L}_{eff}[\varphi,g^{\mu\nu}]}{\partial\dot{\varphi}^2}\bigg\vert_{\psi_i=\mu_i}&=1,&\f{\partial^2\mathcal{L}_{eff}[\varphi,g^{\mu\nu}]}{\partial\varphi^2}\bigg\vert_{\psi_i=\mu_i}&=-m^2,&\f{\partial^4\mathcal{L}_{eff}[\varphi,g^{\mu\nu}]}{\partial\varphi^4}\bigg\vert_{\psi_i=\mu_i}&=-\lambda,\nonumber \\\f{\partial^3\mathcal{L}_{eff}[\varphi,g^{\mu\nu}]}{\partial\varphi^2\partial R}\bigg\vert_{\psi_i=\mu_i}& =-\xi&\mathcal{L}_{eff}[\varphi,g^{\mu\nu}]\big\vert_{\psi_i=\mu_i}& =\Lambda,&\f{\partial\mathcal{L}_{eff}[\varphi,g^{\mu\nu}]}{\partial R}\bigg\vert_{\psi_i=\mu_i}&=\alpha,\nonumber\\\f{\partial^2\mathcal{L}_{eff}[\varphi,g^{\mu\nu}]}{\partial R^2}\bigg\vert_{\psi_i=\mu_i}&=\beta,&\f{\partial\mathcal{L}_{eff}[\varphi,g^{\mu\nu}]}{\partial R_{\alpha\beta}R^{\alpha\beta}}\bigg \vert_{\psi_i=\mu_i}&=\epsilon_{1},&\f{\partial\mathcal{L}_{eff} [\varphi,g^{\mu\nu}]}{\partial R_{\alpha\beta\gamma\delta} R^{\alpha\beta\gamma\delta}} \bigg\vert_{\psi_i
 =\mu_i}&=\epsilon_{2}  \label{eq:renorm1},} 
where $\psi_i$ is a symbolic notation for all the matter \textit{and} gravitational fields\footnote{$\psi_1=\dot{\varphi}$,  $\psi_2={\varphi}$,  $\psi_3=R$,  $\psi_4=R_{\alpha\beta}R^{\alpha\beta}$ and $\psi_5=R_{\alpha\beta\gamma\delta}R^{\alpha\beta\gamma\delta}$.} and where all the scales $\mu_i$ are arbitrary and can be chosen for each constant separately\footnote{Not all scale choices provide analytic solutions. See the discussion at the end of subsection (\ref{sec:eomenergy}).}. This procedure is simple to perform and it allows complete control over the finite parts of each counterterm separately. The effective action approach is however quite cumbersome for deriving the quantum corrected Einstein equation. If, for example, one is interested in the quantum corrections in a Friedmann-Robertson-Walker (FRW) space-time, the effective action must regardless be derived with a general metric and only after renormalization and variation with respect to $g^{\mu\nu}$ as in (\ref{eq:var}) can the space-time be chosen to be homogeneous and isotropic. In general it is quite challenging to derive an expression for the effective action with an arbitrary metric. Usually one must settle for approximations such as the gradient expansion to be described in subsection \ref{sec:derivative}. This has led to the construction of renormalization procedures operating at the level of the equations of motion.

\subsection{Adiabatic subtraction}
\label{sec:subtraction}

Adiabatic subtraction \cite{Parker:1974qw,Fulling:1974zr, Fulling:1974pu} is one of the most popular regularization (renormalization) methods used in curved space-time calculations, in particular for the energy-momentum tensor. The renormalized quantities are defined at the level of the equations of motion by subtracting an $A$th order derivative approximation of the quantity of interest from the divergent expression. These quantities used in subtractions are said to be defined in $A$th order adiabatic vacua to be discussed in subsection \ref{sec:choosing}. The order of the expansion $A$ depends on the adiabatic order of divergences in the bare quantity. Thus in this procedure one only calculates one subtraction term that contains all the divergences needed for rendering a particular expression finite and counterterms such as $\delta m^2$ or $\delta \lambda$ never explicitly enter the picture. For example, the renormalized variance of a field and the quantum part of the energy-momentum tensor defined via adiabatic subtraction read
\ea{\label{eq:adsub1}\langle \hat{\varphi}^2\rangle&=\langle \hat{\varphi}^2\rangle_0+\delta {\varphi}^2\equiv\langle \hat{\varphi}^2\rangle-\langle 0^{(A)}\vert \hat{\varphi}^2\vert 0^{(A)}\rangle\big\vert_{A=2}\\\langle \hat{T}^Q_{\mu\nu}\rangle&=\langle \hat{T}^Q_{\mu\nu}\rangle_0+\delta T_{\mu\nu}\equiv\langle \hat{T}^Q_{\mu\nu}\rangle-\langle 0^{(A)}\vert \hat{T}^Q_{\mu\nu}\vert 0^{(A)}\rangle\big\vert_{A=4}.\label{eq:adsub}}
The argument is that all the constants are taken to be renormalized since the divergences are taken care of by the adiabatic subtraction terms.

This procedure has the advantage that it is generally easier to calculate the energy-momentum tensor starting directly from operator expressions than first deriving a renormalized expression for an effective action and then obtaining the energy-momentum tensor via variation. In particular, at the equation of motion level it is perfectly allowed to constrain the metric to be of special forms e.g. of the FRW-type. This method has its drawbacks, however. Because the renormalized expression is derived by a single subtraction, the renormalization conditions for each constant including the renormalization scale, are not explicit. This leaves the finite part of each counterterm implicit and thus the physical interpretations for the renormalized constants become more difficult. The way to avoid this would be to add additional finite parts into the constants of the action and fixing them according to some chosen conditions, as was done in \cite{Paz:1988mt}. Since this fixing of finite parts ruins the calculational simplicity of adiabatic subtraction, it is rarely used. This is a rather unsatisfactory situation, since ultimately for a renormalizable theory one should be able to perform the entire renormalization process by simply redefining the constants in the classical action. Furthermore, in \cite{Bunch:1980vc} it was shown that for the case $\lambda=0$ in (\ref{eq:actm}), the adiabatic subtraction term in (\ref{eq:adsub}) can be reduced to a redefinition of the bare constants of the action, but it is a non-trivial issue whether this applies also to the interacting theory. This matter will be addressed in subsection \ref{sec:compsubtraction}.
\subsection{Consistent renormalization via the energy-momentum tensor}
\label{sec:eomenergy}
In this section we set out to to find a renormalization procedure that allows control over the finite parts of the counterterms while working at the level of the equations of motion, in particular the Einstein equation. This way one could combine the benefits of the effective action approach and adiabatic subtraction. 

Let us start by defining the energy-momentum tensor. For quantum matter and classical gravity denoted as $S[\hat{\varphi},g^{\mu\nu}]_0\equiv S_g[g^{\mu\nu}]_0+S_m[\hat{\varphi},g^{\mu\nu}]_0$,\footnote{In our notation the subscript $"0"$ denotes a quantity where all the coupling constants are bare constants. For renormalized constants, we simply drop the subscript. A bare constant $c_0$ can be split into a (finite) renormalized part and a counterterm as $c_0=c+\delta c$ } we get the Einstein equation via varying the bare action with respect to the metric
\ee{\label{eq:einE}\f{2}{\sqrt{-g}}\f{\delta S[\hat{\varphi},g^{\mu\nu}]_0}{\delta g^{\mu\nu}}=0\quad\Leftrightarrow\quad\f{2}{\sqrt{-g}}\f{\delta S_g[g^{\mu\nu}]_0}{\delta g^{\mu\nu}}=-\f{2}{\sqrt{-g}}\f{\delta S_m[\hat{\varphi},g^{\mu\nu}]_0}{\delta g^{\mu\nu}}\equiv \hat{T}_{\mu\nu,0}.}
A crucial point is that although $\hat{T}_{\mu\nu,0}$ has all the counterterms coming from the matter part via the bare coupling constants, we must also include counterterms from the gravitational side of (\ref{eq:einE}) in order to remove all the divergences. Indeed, this is precisely the reason for introducing the higher order tensors in (\ref{eq:actg}). We will denote the entire counterterm contributions as
\ee{\delta T_{\mu\nu} \equiv\delta T^{m}_{\mu\nu} -\delta T^{g}_{\mu\nu},}
where $\delta T^{m}_{\mu\nu}$ symbolises the counterterms from the matter action and $\delta T^{g}_{\mu\nu}$ the gravitational action. The renormalized energy-momentum tensor is then

\ee{\label{eq:renomE}\hat{T}_{\mu\nu}=-\f{2}{\sqrt{-g}}\f{\delta S_m[\hat{\varphi},g^{\mu\nu}]}{\delta g^{\mu\nu}}+\delta T_{\mu\nu}.} For the gravitational part in (\ref{eq:actg}) with the help of appendix \ref{sec:appA} we have
\ee{\f{2}{\sqrt{-g}}\f{\delta S_{g}[g^{\mu\nu}]_0}{\delta g^{\mu\nu}}=-g_{\mu\nu}\Lambda_0+2\alpha_0 G_{\mu\nu} +2\beta_0~^{(1)}H_{\mu\nu}+2\epsilon_{1,0}~^{(2)}H_{\mu\nu}+2\epsilon_{2,0}H_{\mu\nu},}
so in addition to the standard counterterms from the matter part of the action, such as $\delta m^2$ and $\delta \lambda$, we have the gravity counterterms
\ee{\delta T^{g}_{\mu\nu}\equiv-g_{\mu\nu}\delta\Lambda+2\delta\alpha G_{\mu\nu} +2\delta\beta~^{(1)}H_{\mu\nu}+2\delta\epsilon_{1}~^{(2)} H_{\mu\nu}+2\delta\epsilon_{2}H_{\mu\nu}.\label{eq:deltaSg}}

We would now like to find a generalization of the equations (\ref{eq:renorm1}) for the energy-momentum tensor, so that we can determine each counterterm in (\ref{eq:renomE}) separately. For this we can use equations very similar to those in (\ref{eq:renorm1}). The only difference being that we use matching between the complete expectation value of the energy-momentum tensor to the classical one for determining the counterterms. 
We no longer need to keep a general metric, since the variation with respect $g^{\mu\nu}$ is already done. Therefore we can choose constraints for the metric, for example that we are in a homogeneous and isotropic space, and use this accordingly in the renormalization conditions.

Now we will derive the renormalization equations for the coupling constants in (\ref{eq:actm}) and (\ref{eq:actg}) in a space-time with a FRW-type metric $g_{\mu\nu}dx^\mu dx^\nu=-dt^2+a(t)d\mathbf{x}^2$ with a non-zero mass and a possible time dependence in the field $\varphi$. 
Since we will employ dimensional regularization in section \ref{sec:expli}, these formulae are calculated in $n$ dimensions. Using appendix \ref{sec:appA} we straightforwardly derive the result for the classical energy-momentum tensor
\ee{T_{00}^C=\f{1}{2}\bigg[\dot{\varphi}^2+m^2\varphi^2+2\f{\lambda}{4!}\varphi^4 \bigg]+\xi\bigg[(n-1)\bigg(\f{n}{2}-1\bigg)\bigg(\f{\da}{a}\bigg)^2\varphi^2+2(n-1)\f{\da}{a}\dot{\varphi}\varphi\bigg],}
which gives the renormalization conditions\footnote{Depending on ones interests, it is also possible to include the contribution of the entire classical energy-density on the right hand side of each condition in order to get a different finite part for the renomalization constant. For example for the mass term one could then write the condition  $ \partial^2 \langle\hat{T}_{00}\rangle/\partial\varphi^2\big\vert_{\psi_i=\mu_i}=m^2+\f{\lambda}{2}\varphi^2_0 $ }
\ea{\f{\partial^2\langle\hat{T}_{00}\rangle}{\partial\dot{\varphi}^2} \bigg\vert_{\psi_i=\mu_i}&=1,&\f{\partial^2 \langle\hat{T}_{00}\rangle}{\partial\varphi^2}\bigg\vert_{\psi_i=\mu_i}&=m^2, &\f{\partial^4\langle\hat{T}_{00}\rangle}{\partial\varphi^4}\bigg\vert_{\psi_i=\mu_i}&=\lambda,\nonumber \\ \f{\partial^3\langle\hat{T}_{00}\rangle}{\partial\varphi\partial\dot{\varphi} \partial(\da/a)}\bigg\vert_{\psi_i=\mu_i}&={2\xi(n-1)}&\langle\hat{T}_{00} \rangle\big\vert_{\psi_i=\mu_i}&=0, &\f{\partial^2 \langle\hat{T}_{00}\rangle}{\partial (\da/a)^2}\bigg\vert_{\psi_i=\mu_i}&=0\nonumber \\\f{\partial^3 \langle\hat{T}_{00}\rangle}{\partial(\da/a)^2\partial (\dda/a)}\bigg\vert_{\psi_i=\mu_i}&=0,&\f{\partial^2 \langle\hat{T}_{00}\rangle}{\partial(\dda/a)^2}\bigg\vert_{\psi_i=\mu_i}&=0,&\f{\partial^4 \langle\hat{T}_{00}\rangle}{\partial(\da/a)^4}\bigg\vert_{\psi_i=\mu_i}&=0.\label{eq:renormT2}}
Here again $\varphi$ symbolizes the expectation value of the field just like in (\ref{eq:renorm1}) and the scale choices for each constant can be done independently. The first four conditions in the above fix the counterterms coming from the matter part of the action and the rest do the same for the counterterms from the gravity part. It is obvious that there are many other possible choices for renormalization equations. For example $\delta\xi$ can also be determined from
\ee{\label{eq:dcond}\f{\partial^4\langle\hat{T}_{00}\rangle}{\partial\varphi^2\partial (\da/a)^2}\bigg\vert_{\psi_i=\mu_i}=4\xi (n-1)\bigg(\f{n}{2}-1\bigg).}
Also, we could have chosen any other component of $T_{\mu\nu}$ for determining the renormalization constants. Had we chosen to use different conditions to those in (\ref{eq:renormT2}), the resulting counterterms potentially would have had different finite parts.

A few comments are now in order. First, since renormalization was achieved by simply including counterterms in the constants of the original action, covariant conservation is automatically satisfied\footnote{This can be shown by operating with $\nabla^\mu$ on the gravitational side of (\ref{eq:einE}) and using the Bianchi identities and commutator formulae.}. If the divergences are removed by introducing a subtraction term by hand, covariant conservation must be separately checked. Second, a crucial requirement for the equations (\ref{eq:renormT2}) to work is that all the equations are analytic at the chosen scales $\mu_i$. As was discussed in \cite{Coleman:1973jx} in the $\varphi^4$ theory there is an infrared singularity in the massless limit. To bypass this issue at the renormalization stage, one may choose non-zero renormalization scales $\mu_i$. However, changing renormalization scales changes the definitions of the constants and ultimately changes the parameter space where the perturbative expansion is valid. An example of using a non-zero renormalization scale can be found in subsection \ref{sec:confa}. To cure the IR problem altogether requires further resummations \cite{Serreau:2011fu}.

\section{Field equations and the energy-momentum tensor}
\label{sec:Field equations}


\subsection{Choosing the vacuum}
\label{sec:choosing} 

For our calculations we will use the adiabatic vacuum\footnote{This is just one of many possible choices of vacuum, but one that allows direct comparison with adiabatic subtraction and the Schwinger-DeWitt expansion.} \cite{ParkerToms} and we must perform our calculation in $n$ dimensions since dimensional regularization is to be used in section \ref{sec:expli}.
With a field equation of motion of the form
\ee{\label{eq:eom}\bigg[-\square+M^2(t)\bigg]{\hat\varphi}=0,} 
where $M(t)$ is some arbitrary and possibly time-dependent mass parameter, we can solve it in terms of mode functions by using an ansatz \eg{\hat{\varphi}=\int d^{n-1}k\big [{a}_\mathbf{k} u_\mathbf{k}+{a}^*_\mathbf{k} u_\mathbf{k}^*\big],\quad u_\mathbf{k}=\f{1}{\sqrt{2(2\pi)^{n-1}a^{n-1}}}h_\mathbf{k}(t)e^{i\mathbf{k}\cdot \mathbf{x}},\nonumber \\h_\mathbf{k}(t)=\f{1}{\sqrt{W}}e^{-i\int^{t}Wdt'}\label{eq:ans}} 
and assuming $g_{\mu\nu}dx^\mu dx^\nu=-dt^2+a(t)d\mathbf{x}^2$, we can write $W$ as an adiabatic expansion
\ee{W=c_0+c_1\f{\da}{a}+c_2\f{\dot{M}}{M}+c_3\f{\da^2}{a^2}+c_4\f{\dot{M}^2}{M^2}+c_5\f{\dda}{a}+c_6\f{\ddot{M}}{M}+c_7\f{\da\dot{M}}{aM}+\cdots,} with $c_i$ being functions of $M$ and $a$.
This approximation can be trusted when $\varphi$ and $a$ are slowly varying. The solution for $u_k$ will naturally also be an expansion in the number of time derivatives and the $A$th order approximate solution will include all terms with an $A$ number of derivatives\footnote{This means that $\da^2$ and $\dda$ are of the same adiabatic order, for example.} and it will be denoted as $u_\mathbf{k}^{(A)}$. Similarly, the vacuum it defines is written as $\vert 0^{(A)}\rangle$. These approximate modes can be used to define an exact solution to the equation of motion (\ref{eq:eom}) through the relation
\ee{u_\mathbf{k}=\alpha_{\bf k}(t)u^{(A)}_\mathbf{k}+\beta_{\bf k}(t)u^{(A)*}_\mathbf{k},}
where $\alpha_{\bf k}$ and $\beta_{\bf k}$ must be constant in $t$ to order $A$, but they may have dependence in $\mathbf{k}$. We can then choose to fix the exact mode $u_\mathbf{k}$ at some point $t=t_0$ to be the $A$th order positive solution:
\ee{\alpha_{\bf k}(t_0)=1+\mathcal{O}(A+1)\quad \beta_{\bf k}(t_0)=0+\mathcal{O}(A+1)}
This defines the $A$th order adiabatic vacuum. Clearly this procedure has some ambiguity, since we can choose an arbitrary point $t_0$ as our condition to fix the exact mode resulting in different $A$th order vacua $u_\mathbf{k}$. However, the exact modes differ only in the $A+1$ adiabatic order, so if we choose to neglect $\mathcal{O}(A+1)$ we can write the quantized field operator to $\mathcal{O}(A)$ as
\ee{\hat{\varphi}(x)=\int{dk^{n-1}}\big[\hat{a}_\mathbf{k}u_\mathbf{k}^{(A)}+ \hat{a}_\mathbf{k}^{\dagger}u_\mathbf{k}^{(A)*}\big],}
with the standard commutation relations
\ee{[\hat{a}_\mathbf{k},\hat{a}_{\mathbf{k}'}]=[{\hat{a}}^\dagger_\mathbf{k},{\hat{a}}^\dagger_{\mathbf{k}'}]=0,\quad[{\hat{a}}_\mathbf{k},{\hat{a}}^\dagger_{\mathbf{k}'}]=\delta(\mathbf{k}-\mathbf{k}'),}
and remember that in principle it is the exact mode that becomes quantized and defines the vacuum. Troughout this paper we will work in the adiabatic approximation, where we include only two or four time derivatives ($A=2$ or $4$).

\subsection{Field equation to one-loop order}
\label{sec:mode}
From the action in (\ref{eq:actm}) the equation of motion for the operator $\hat{\varphi}$ reads
\ee{
\bigg[-\Box +m^2_0+\xi_0 R\bigg]\hat{\varphi}+\f{\lambda_0}{3!}\hat{\varphi}^3=0.
\label{eq:eom1}}
Since we are interested in the one-loop corrections, we can shift the field operator as $\hat{\varphi}\rightarrow\varphi+\hat{\phi}$, where $\langle\hat{\varphi}\rangle\equiv\varphi$ as in section \ref{sec:effective}, and expand (\ref{eq:actm}) around $\hat{\phi}=0$ giving to quadratic order\footnote{In the one-loop approximation, the terms linear in $\hat{\phi}$ can be discarded. This is because we can always choose a renormalization condition for a linear term in the classical action that cancels this contribution \cite{Peskin:1995ev}.}
\ea{\label{eq:expS}S_m[\varphi,\hat{\phi},g^{\mu\nu}]_0=&-\f{1}{2}\int d ^nx\sqrt{-g}~ \bigg[\partial_\mu\varphi\partial^\mu\varphi+m^2_0\varphi^2 +\xi_0 R\varphi^2+2\f{\lambda_0}{4!}\varphi^4\bigg]\nonumber \\ &-\f{1}{2}\int d ^nx\sqrt{-g}~\hat{\phi}\bigg[-\Box +m^2_0+\xi_0 R+\f{\lambda_0}{2}\varphi^2\bigg]\hat{\phi}+\cdots}
In a similar fashion we can also split (\ref{eq:eom1}) into two coupled equations, one for the expectation value of the original mean field and one for the shifted operator
\ea{&\bigg[-\Box +m^2_0+\xi_0 R\bigg]\varphi+\f{\lambda_0}{3!}\varphi^3+\f{\lambda_0}{2}\varphi\langle\hat{\phi}^2\rangle=0\label{eq:eom2}\\ &\bigg[-\Box +m^2_0+\xi_0 R+\f{\lambda_0}{2}\varphi^2\bigg]\hat{\phi}=0.\label{eq:eom3}
}
Note that we also used the fact that since $\hat{\phi}$ is Gaussian at one loop order, all correlators with an odd number of fields vanish. Again, as indicated by the subscripts $"0"$, we need to split our bare coupling into a renormalized part and a counterterm. Since we include only the first non-trivial correction and since the counterterms are the result of the one-loop quantum correction, all terms such as $\delta\lambda\langle\phi^2\rangle$ are effectively two-loop terms and beyond our approximation\footnote{This amounts to counting loops at the level of the effective action, rather than at the level of the equations of motion. The discarded term would be included if the the two-loop "figure-8" diagram was included at the level of the action.}. Inserting the counterterms into equations (\ref{eq:eom2}) and (\ref{eq:eom3}) we get 
\ea{&\bigg[-\Box +m^2+\delta m^2+(\xi+\delta\xi) R\bigg]\varphi+\f{\lambda+\delta\lambda}{3!}\varphi^3+\f{\lambda}{2}\varphi\langle\hat{\phi}^2\rangle=0\label{eq:eomR2}\\ &\bigg[-\Box +m^2+\xi R+\f{\lambda}{2}\varphi^2\bigg]\hat{\phi}=0.\label{eq:eomR3}
}
In the above we used the fact that in a one-loop calculation there is no need for a wave function counterterm multiplying the d'Alembertian operator\footnote{Even if there is no divergence proportional to the kinetic term, one might still need a counterterm in order to give a precise physical interpretation for this quantity. For our renormalization scale choices this will be of no relevance.}.

The equation for the operator (\ref{eq:eom3}) $\hat{\phi}$
\ee{ \Leftrightarrow\quad\bigg[\partial_t\partial_t +(n-1)\f{\da}{a}\partial_t-a^{-2}\partial_i\partial^i+m^2+\f{\lambda}{2}\varphi^2+\xi R\bigg]\hat{\phi}=0,
}
is precisely of the form (\ref{eq:eom}) and hence we can use the ansatz from (\ref{eq:ans}), which after some algebra gives an equation for $W$
\ee{\label{eq:W}W^2=\sigma(t)+\f{3\dot{W}^2}{4W^2}-\f{\ddot{W}}{2W},}
where
\ee{
\sigma(t)=\underbrace{\mathbf{k}^2/a^2+m^2+\f{\lambda}{2}\varphi^2}_{\displaystyle \equiv\omega^2}+\f{\dda}{a}\bigg[\f{1}{2}(n-1)(4\xi-1)\bigg]+\bigg(\f{\da}{a}\bigg)^2\bigg[\f{1}{4}(n-1)\big(3-n+4(n-2)\xi\big)\bigg]
}
and the $n$ dimensional scalar curvature in FRW was found from (\ref{eq:Rt}). $W^2$ can now be solved from (\ref{eq:W}) iteratively as
\ee{\label{eq:Wexp}W^2=(W^2)^{(0)}+(W^2)^{(1)}+(W^2)^{(2)}+(W^2)^{(3)}+\cdots,}
where, as explained in section \ref{sec:choosing}, the expansion parameter is the number of derivatives. We will need $W$ only up to four derivatives and the complete results are relegated to appendix \ref{sec:Sol}.


\subsection{Energy-momentum tensor to one-loop order}
\label{sec:energy}

From (\ref{eq:expS}) we get the $n$-dimensional unrenormalized energy-momentum tensor to one-loop order by straightforward variation:
\eg{\f{-2}{\sqrt{-g}}\f{\delta S_m[\varphi,\hat{\phi},g^{\mu\nu}]_0}{\delta g^{\mu\nu}}=\hat{T}_{\mu\nu,0}\nonumber \\=-\f{g_{\mu\nu}}{2}\bigg[\partial_\rho\varphi\partial^\rho\varphi+m^2_0\varphi^2+2\f{\lambda_0}{4!}\varphi^4 \bigg]+\partial_\mu\varphi\partial_\nu\varphi +\xi_0\big[G_{\mu\nu}-\nabla_\mu\nabla_\nu+g_{\mu\nu}\Box\big]\varphi^2\nonumber \\-\f{g_{\mu\nu}}{2}\bigg[\partial_\rho\hat{\phi}\partial^\rho\hat{\phi} +m^2_0\hat{\phi}^2+\f{\lambda_0}{2}\varphi^2\hat{\phi}^2 \bigg]+\partial_\mu\hat{\phi}\partial_\nu\hat{\phi}+ \xi_0\big[G_{\mu\nu}-\nabla_\mu\nabla_\nu+g_{\mu\nu}\Box\big]\hat{\phi}^2\nonumber \\ \equiv T^C_{\mu\nu,0}+\hat{T}^Q_{\mu\nu}.}
That the energy-momentum tensor nicely splits into a classical and a quantum part is slightly misleading since only the complete $\hat{T}_{\mu\nu,0}$ has the right conservation properties. Including the counterterms from the gravity side and separating $T^C_{\mu\nu,0}$ into renormalized and counterterm parts, we have for the renormalized E-M tensor
\eg{{\hat{T}_{\mu\nu}}=T^C_{\mu\nu}+\hat{T}^Q_{\mu\nu}+\delta T_{\mu\nu},\label{eq:RT}}
where the counterterm contributions for our one-loop approximation are
\ee{\delta T_{\mu\nu}=-\f{g_{\mu\nu}}{2}\bigg[\delta m^2\varphi^2+2\f{\delta\lambda}{4!}\varphi^4 \bigg]+\delta\xi\big[G_{\mu\nu}-\nabla_\mu\nabla_\nu+g_{\mu\nu}\Box\big]\varphi^2-\delta T^{g}_{\mu\nu},\label{eq:renT}}
where $\delta T^{g}_{\mu\nu}$ was defined in (\ref{eq:deltaSg}).

In the four dimensional conformal limit we have $m=0$ and $\xi=1/6$. For our $n$ dimensional expression we then have \ee{m_0^2 =0\quad \text{and}\quad \xi_0=\f{n-2}{4(n-1)}+\delta\xi.} Inserting these into $T^C_{\mu\nu}$ and $\langle\hat{T}^Q_{\mu\nu}\rangle$, we get the relations
\ee{g^{\mu\nu}{T^C_{\mu\nu,0}}=\varphi^2\f{n-2}{2}\f{\lambda}{2}\langle\phi^2\rangle+(n-4)\frac{\lambda+\delta\lambda}{4!}\varphi^4 +\delta\xi(n-1)\Box\varphi^2,\quad{g^{\mu\nu}\langle \hat{T}^{Q}_{\mu\nu}}\rangle=-\varphi^2\f{\lambda}{2}\langle\hat{\phi}^2\rangle,} giving for the full trace
\ee{\label{eq:trace}{\langle\hat{T}_\mu^{~\mu}\rangle}=(n-4)\bigg[\f{\lambda\varphi^2}{4}\langle\hat{\phi}^2\rangle+\frac{\lambda+\delta\lambda}{4!}\varphi^4\bigg]+\delta\xi(n-1)\Box\varphi^2-\delta {{T^{g}}_\mu}^\mu,}
which of course will result in the famous conformal anomaly
first seen in \cite{Capper:1974ic} (for general results, see \cite{Deser:1976yx} and \cite{Deser:1993yx}). The standard purely gravitational contribution is from the last term of (\ref{eq:trace}) and the first three give the one-loop $\lambda$-dependent terms, which are studied in more detail in \cite{Drummond:1977dg,Hathrell:1981zb}. The equations in this section also provide a very non-trivial check for the solutions to be obtained in section \ref{sec:expli}.


\section{Explicit results}
\label{sec:expli}

Next we will proceed to use the methods described in previous sections for the theory defined by (\ref{eq:actm}) and (\ref{eq:actg}) in the four dimensional case and present the result for the equation of motion for the field $\varphi$ and the expression for the energy-momentum tensor calculated in the adiabatic vacuum as defined in subsection \ref{sec:choosing}. A point worth emphasizing is that because all the calculations are done in $n$ dimensions, everything is dimensionally regularized and hence, in practice, finite\footnote{In principle one should arrive at the same expressions using a simple cut-off regularisation, when performed with care. But since this breaks general covariance, additional counterterms need to be included, making the calculation more cumbersome \cite{Hollenstein:2011cz}.}. 
For consistency it is important that one starts from an $n$-dimensional Lagrangian and calculates all the required quantities in arbitrary dimensions (as we have done), as opposed to working down from a 4-dimensional theory and only regularizing explicitly divergent equations. We end the section by calculating the result for the anomalous four dimensional trace. We will notate the classical parts and the various adiabatic orders of the quantum contributions as superscripts  \ea{\langle\hat{T}_{\mu\nu}\rangle&=T^C_{\mu\nu}+\langle\hat{T}^Q_{00}\rangle^{(0)} +\langle\hat{T}^Q_{00}\rangle^{(2)}+ \langle\hat{T}^Q_{00}\rangle^{(4)}\nonumber \\ \bigg\langle\f{\delta S[\hat{\varphi},g^{\mu\nu}]_0}{\delta\hat{\varphi}(x)}\bigg\rangle&=\mathcal{E}^C+\mathcal{E}^{(0)}+\mathcal{E}^{(2)}+\mathcal{E}^{(4)}=0.}
\label{sec:explicit}


\subsection{Results to fourth adiabatic order}
\label{sec:second}
Using the expressions from appendix \ref{sec:appA} and subsection \ref{sec:choosing}, we can write the non-trivial components of the quantum energy-momentum tensor in the adiabatic vacuum as 
\ee{ \langle\hat{T}^Q_{00}\rangle =\int d^{n-1} {k}\bigg\{\f{1}{2}\bigg[\vert\dot{u}_\mathbf{k}\vert^2+\bigg( {\mathbf{k}^2/a^2+M^2}\bigg)\vert u_\mathbf{k}\vert^2\bigg]+\xi\bigg[G_{00}+(n-1)\f{\da}{a}\partial_0\bigg]\vert u_\mathbf{k}\vert^2\bigg\}\label{eq:T00}}
and
\ee{
\langle\hat{T}^Q_{ii} \rangle=\int d^{n-1} k\bigg\{\f{a^2}{2}\bigg[\vert\dot{u}_\mathbf{k}\vert^2-\bigg(\f{3-n}{1-n}\f{\mathbf{k}^2}{a^2}+M^2\bigg)\vert u_\mathbf{k}\vert^2\bigg]+\xi\bigg[G_{ii}+a^2\bigg((2-n)\f{\da}{a}\partial_0-\partial_0^2\bigg)\bigg]\vert u_\mathbf{k}\vert^2\bigg\}\label{eq:Tii},}
where $M^2\equiv m^2+\f{\lambda}{2}\varphi^2$. By inserting (\ref{eq:W0} - \ref{eq:W4}) into (\ref{eq:T00}) gives the explicit result for "00" components of the quantum part of the energy-momentum tensor, which 
can be used to perform renormalization as explained in subsection \ref{sec:eomenergy} with the scale choices
\eg{\f{\partial^2\langle\hat{T}_{00}\rangle}{\partial\varphi^2}\bigg\vert_{\varphi=0,a=1}=m^2,\quad\f{\partial^4\langle\hat{T}_{00}\rangle}{\partial\varphi^4}\bigg\vert_{\varphi=0,a=1}=\lambda,\quad\f{\partial^3\langle\hat{T}_{00}\rangle}{\partial\varphi\partial\dot{\varphi}\partial(\da/a)}\bigg\vert_{\varphi=0,a=1}={2\xi(n-1)},\nonumber \\ \langle\hat{T}_{00}\rangle\big\vert_{\varphi=0,a=1}=0,\quad\f{\partial^2 \langle\hat{T}_{00}\rangle}{\partial (\da/a)^2}\bigg\vert_{\varphi=0,a=1}=0\quad\f{\partial^3 \langle\hat{T}_{00}\rangle}{\partial(\da/a)^2\partial (\dda/a)}\bigg\vert_{\varphi=0,a=1}=0,\nonumber \\\f{\partial^2 \langle\hat{T}_{00}\rangle}{\partial(\dda/a)^2}\bigg\vert_{\varphi=0,a=1}=0,\quad\f{\partial^4 \langle\hat{T}_{00}\rangle}{\partial(\da/a)^4}\bigg\vert_{\varphi=0,a=1}=0\label{cond},}
where the condition $a=1$ states that all derivatives of $a$ are zero. 

This choice of renormalising the cosmological constant to zero in a Minkowski space is consistent, but ignores the late-time cosmological acceleration of the present Universe. If one wishes to use renormalization conditions with a space-time not close to Minkowski, the calculation for the vacuum mode in section \ref{sec:choosing} must be done with boundary conditions appropriate for the particular space-time. The subtleties and technical challenges of the more exact renormalization condition in a self-consistent way including this cosmological constant have been considered in \cite{shapiro2009} (and references therein). However, for the physics far from the IR scale of the cosmological constant, we expect the Minkowski vacuum to be a very good approximation.

By using (\ref{eq:counter1} - \ref{eq:cond2}) in the expression for the full $\hat{T}_{\mu\nu}$ in (\ref{eq:RT}) calculated with (\ref{eq:W0} - \ref{eq:W4}) and using standard formulae for $n$-dimensional integrals \cite{Peskin:1995ev},
for $n=4$ we get
\ea{\label{eq:EOMRR1} &T^C_{00}+\langle\hat{T}^Q_{00}\rangle^{(0)}+\langle\hat{T}^Q_{00}\rangle^{(2)}=\f{1}{2}\dot{\varphi}^2+\f{1}{2}m^2\varphi^2+\f{\lambda}{4!}\varphi^4+\xi\bigg(3\f{\varphi^2\da^2}{a^2}+6\f{\varphi\dot{\varphi}\da}{a}\bigg)\nonumber \\&+\f{1}{64\pi^2}\bigg\{(1-6\xi)\lambda\varphi^2\f{\da^2}{a^2}-\f{3}{8}\lambda^2\varphi^4-\f{\lambda^2m^2\varphi^4}{4M^4}+\f{\lambda\varphi^2\big(-3m^4+\lambda\dot{\varphi}^2\big)}{6M^2}\nonumber \\&+\log\bigg(\f{M^2}{m^2}\bigg)\bigg[M^4-2(1-6\xi)M^2\f{\da^2}{a^2}-(1-6\xi)2\lambda\f{\varphi\dot{\varphi}\da}{a}\bigg]\bigg\}}
and \ea{\label{eq:EOMRRR1}\langle\hat{T}^Q_{00}\rangle^{(4)}&=\f{1}{64\pi^2}\bigg\{(1-6 \xi )^2 \left(-\frac{\dda^2}{a^2}-3\frac{\da^4}{a^4}+2\frac{ a^{(3)} \da}{a^2}+2\frac{ \da^2 \dda}{a^3}\right) \log \left(\frac{M^2}{m^2}\right)\nonumber \\&+\bigg[\frac{2}{15} \lambda  \left(540 \xi ^2-255 \xi +29\right) \varphi\bigg]\f{ \dot{a}^3 \dot{\varphi } }{ a^3 M^2}+\bigg[\frac{2}{5} \lambda  (1-5 \xi ) \varphi\bigg]\frac{\ddot{a} \ddot{\varphi }}{a M^2}+\bigg[\frac{\lambda ^2 \varphi ^2}{60}\bigg]\f{\ddot{\varphi}^2}{M^4}\nonumber \\&+\bigg[\frac{1}{15} \lambda ^2 \varphi  \left(\lambda  \varphi ^2+12 m^2 (1-5 \xi )\right)\bigg]\frac{\dot{a} \dot{\varphi }^3}{a M^6}+\bigg[\frac{2}{5} \lambda  \left(180 \xi ^2-55 \xi +4\right) \varphi\bigg]\frac{\dot{a} \dot{\varphi } \ddot{a}}{a^2 M^2}\nonumber \\&+ \bigg[\frac{1}{10} \lambda  \left(\varphi ^2 (\lambda -10 \lambda  \xi )+4 m^2 (1-5 \xi )\right)\bigg]\frac{\dot{\varphi }^2 \ddot{a}}{a M^4}+\bigg[\frac{6}{5} \lambda  (5 \xi -1) \varphi\bigg]\frac{\dot{a}^2 \ddot{\varphi }}{a^2 M^2}\nonumber \\&+\bigg[\frac{1}{5} \lambda  (5 \xi -1) \left(6 m^2-\lambda  \varphi ^2\right)\bigg]\frac{\dot{a} \dot{\varphi } \ddot{\varphi }}{a M^4}+\bigg[\frac{2}{5} \lambda  (5 \xi -1) \varphi\bigg]\frac{\dot{a} \varphi ^{(3)}}{a M^2}+\bigg[-\frac{1}{30} \lambda ^2 \varphi ^2\bigg]\frac{\dot{\varphi } \varphi ^{(3)}}{M^4}\nonumber \\&+\bigg[\frac{1}{240} \lambda ^2 \left(-\lambda ^2 \varphi ^4+4 m^4+20 \lambda  m^2 \varphi ^2\right)\bigg]\frac{\dot{\varphi }^4}{M^8}+\bigg[\frac{2}{5} \lambda  (5 \xi -1) \varphi\bigg]\frac{a^{(3)} \dot{\varphi }}{a M^2}\nonumber \\&+\bigg[\frac{1}{20} \lambda  \left(\lambda  (19-80 \xi ) \varphi ^2+24 m^2 (5 \xi -1)\right)\bigg]\frac{\dot{a}^2 \dot{\varphi }^2}{a^2 M^4}+\bigg[\frac{1}{30} \lambda ^2 \varphi  \left(\lambda  \varphi ^2-2 m^2\right)\bigg]\frac{\dot{\varphi }^2 \ddot{\varphi }}{M^6}\bigg\}.}
The same counterterms renormalize the "$ii$" component of the energy-momentum tensor, with $\hat{T}^Q_{ii}$ given by (\ref{eq:Tii}), and the equation of motion for the field $\varphi$ (\ref{eq:eomR2}). Similar steps that led to (\ref{eq:EOMRR1}) give
\ea{\label{eq:EOMRR2}&T^C_{ii}+\langle\hat{T}^Q_{ii}\rangle^{(0)}+\langle\hat{T}^Q_{ii}\rangle^{(2)}\nonumber \\&=a^2\bigg(\f{1}{2}\dot{\varphi}^2-\f{1}{2}m^2\varphi^2-\f{\lambda}{4!}\varphi^4\bigg)- a^2\xi\bigg[\varphi^2\bigg(\f{\da^2}{a^2}+2\f{\dda}{a}\bigg)+4\varphi\dot{\varphi}\f{\da}{a}+2\dot{\varphi}^2+2\varphi\ddot{\varphi}\bigg]\nonumber \\&+\f{a^2}{64\pi^2}\bigg\{-\f{\lambda(1-6\xi)\varphi^2}{3}\bigg(\f{\da^2}{a^2}+\f{2\dda}{a}\bigg)+\f{\lambda\varphi^2\big(3m^4+\lambda(5-24\xi)\dot{\varphi}^2\big)}{6M^2}+\f{\lambda^2\varphi^4}{24}\bigg(9+\f{6m^2}{M^2}\bigg)\nonumber \\&+\log\bigg(\f{M^2}{m^2}\bigg)\bigg[-M^4+(1-6\xi)\bigg(\f{4}{3}{\lambda\varphi\dot{\varphi}}\f{\da}{a}+\f{2}{3}M^2\bigg(\f{\da^2}{a^2}+2\f{\dda}{a}\bigg)+\f{2}{3}\lambda(\dot{\varphi}^2+\varphi\ddot{\varphi})\bigg)\bigg]\bigg\},}

\ea{\label{eq:EOMRRR2}\langle\hat{T}^Q_{ii}\rangle^{(4)}&=\f{a^2}{64\pi^2}\bigg\{-(1-6 \xi )^2 \left(\frac{2 a^{(4)}}{3 a}+\frac{\dda^2}{a^2}+\frac{\da^4}{a^4}+\frac{4 a^{(3)} \da}{3 a^2}-\frac{4 \da^2 \dda}{a^3}\right) \log \left(\frac{M^2}{m^2}\right)\nonumber \\&+\bigg[-\frac{1}{36} \lambda ^2 (36 (5-12 \xi ) \xi -17) \varphi ^2-\frac{4}{9} \lambda  m^2 \left(54 \xi ^2-21 \xi +2\right)\bigg]\frac{\dot{a}^2 \dot{\varphi }^2}{a^2 M^4}\nonumber \\&+\bigg[-\frac{1}{18} \lambda ^3 (18 \xi -5) \varphi ^3-\frac{1}{9} \lambda ^2 m^2 (17-78 \xi ) \varphi\bigg]\frac{\dot{a} \dot{\varphi }^3}{a M^6}+\bigg[-\frac{2}{3} \lambda  \xi  (36 \xi -7) \varphi\bigg]\frac{\ddot{a} \ddot{\varphi }}{a M^2}\nonumber \\&+\bigg[\frac{4\lambda  (1-6 \xi )^2 \varphi  }{3}\bigg]\f{ \dot{a}^3 \dot{\varphi } }{ a^3 M^2}+\bigg[\frac{2}{9} \lambda  (9 \xi  (19-48 \xi)-17) \varphi\bigg]\frac{\dot{a} \dot{\varphi } \ddot{a}}{a^2 M^2}+\bigg[\frac{2}{15} \lambda  (1-5 \xi )\varphi \bigg]\frac{ \varphi ^{(4)}
 }{M^2}\nonumber \\&+ \bigg[\lambda\left(\frac{1}{18} \lambda  (54 (4 \xi -1) \xi +1) \varphi ^2-\frac{2}{3} m^2 \xi  (36 \xi -7)\right)\bigg]\frac{\dot{\varphi }^2 \ddot{a}}{a M^4}+\bigg[\frac{4}{3} \lambda  (6 \xi -1)^2 \varphi\bigg]\frac{a^{(3)} \dot{\varphi }}{a M^2}\nonumber \\&+\bigg[\frac{1}{9} \lambda  (12 \xi  (7-18 \xi )-8) \varphi\bigg]\frac{\dot{a}^2 \ddot{\varphi }}{a^2 M^2}+\bigg[\frac{1}{9} \lambda  \left(\lambda  (33 \xi -8) \varphi ^2+m^2 (18-90 \xi )\right)\bigg]\frac{\dot{a} \dot{\varphi } \ddot{\varphi }}{a M^4}\nonumber \\&+\bigg[-\lambda ^2 \left(\frac{\lambda ^2 \varphi ^4}{144}+\frac{1}{60} m^4 (17-80 \xi )+\frac{1}{180} \lambda  m^2 (600 \xi -137) \varphi ^2\right)\bigg]\frac{\dot{\varphi }^4}{M^8}\nonumber \\&+\bigg[-\lambda ^2 \varphi  \left(\frac{1}{9} \lambda  (3 \xi -1) \varphi ^2+\frac{1}{15} m^2 (28-130 \xi )\right)\bigg]\frac{\dot{\varphi }^2 \ddot{\varphi }}{M^6}\nonumber \\&+\bigg[-\frac{2}{3} \lambda  (5 \xi -1) \varphi\bigg]\frac{\dot{a} 
 \varphi ^{(3)}}{a M^2}+\bigg[\frac{1}{30} \left(-5 \lambda ^2 (1-4 \xi ) \varphi ^2-16 \lambda  m^2 (5 \xi -1)\right)\bigg]\frac{\dot{\varphi } \varphi ^{(3)}}{M^4}\nonumber \\&+\bigg[\frac{1}{5} \lambda  m^2 (2-10 \xi )-\frac{1}{12} \lambda ^2 (1-4 \xi ) \varphi ^2\bigg]\f{\ddot{\varphi}^2}{M^4}\bigg\}}
and for the field equation of motion (\ref{eq:eom2})
\ea{\label{eq:EOMRR3}&\mathcal{E}^C+\mathcal{E}^{(0)}+\mathcal{E}^{(2)}=\ddot{\varphi}+3\f{\da}{a}\dot{\varphi}+6\xi \varphi\bigg(\f{\da^2}{a^2}+\f{\dda}{a}\bigg) +m^2\varphi+\f{\lambda}{6}\varphi^3+\f{1}{64\pi^2}\bigg\{-\lambda^2\varphi^3+\f{\lambda^2\varphi^2\dot{\varphi}}{M^2}\f{\da}{a}\nonumber \\&+\f{m^2\lambda^2\varphi\dot{\varphi}^2}{3M^4}+\f{\lambda^2\varphi^2\ddot{\varphi}}{3M^2}+2\lambda\bigg[\varphi M^2-(1-6\xi)\varphi \bigg(\f{\da^2}{a^2}+\f{\dda}{a}\bigg)\bigg]\log\bigg(\f{M^2}{m^2}\bigg)\bigg\}}
and
\ea{\label{eq:EOMRRR3}\mathcal{E}^{(4)}&=\f{1}{64\pi^2}\bigg\{\bigg[\lambda  (1-6 \xi )^2 \varphi\bigg]\f{ \dot{a}^4  }{ a^4M^2}+\bigg[-\frac{1}{30} \lambda ^2 \varphi ^2\bigg]\frac{\varphi ^{(4)}}{M^4}+\bigg[\frac{1}{20} \lambda ^2 \varphi  \left(\lambda  \varphi ^2-2 m^2\right)\bigg]\f{\ddot{\varphi}}{M^6}\nonumber \\&+\bigg[\frac{2}{15} \lambda  \left(540 \xi ^2-255 \xi +29\right) \varphi\bigg]\frac{\dot{a}^2 \ddot{a}^2}{a^3 M^2}+\bigg[\frac{6}{5} \lambda  (5 \xi -1) \varphi\bigg]\frac{\dot{a} a^{(3)}}{a^2 M^2}+\bigg[\frac{2}{5} \lambda  (5 \xi -1) \varphi\bigg]\frac{{a}^{(4)}} {a M^2}\nonumber \\&+ \bigg[\frac{1}{5} \lambda  \left(180 \xi ^2-50 \xi +3\right) \varphi\bigg]\frac{\ddot{a}^2}{a^2 M^2}+\bigg[-\frac{1}{30} \lambda ^2 (60 \xi -13) \varphi ^2\bigg]\frac{\dot{a}^3 \dot{\varphi }}{a^3 M^4}\nonumber \\&+\bigg[\frac{1}{60} \lambda ^2 \varphi  \left(\varphi ^2 (60 \lambda  \xi +\lambda )+2 m^2 (7-60 \xi )\right)\bigg]\frac{\dot{a}^2 \dot{\varphi }^2}{a^2 M^6}+\bigg[\frac{1}{15} \lambda ^3 \varphi ^2 \left(11 m^2-2 \lambda  \varphi ^2\right)\bigg]\frac{\dot{\varphi }^3\da}{aM^8}\nonumber \\&+\bigg[-\frac{2}{15} \lambda ^2 (60 \xi -11) \varphi ^2\bigg]\frac{\da\ddot{a }\dot{\varphi} }{a^2M^4}+\bigg[\frac{1}{30} \lambda ^2 \varphi  \left(\lambda  (30 \xi -1) \varphi ^2+6 m^2 (1-10 \xi )\right)\bigg]\f{\ddot{a}\dot{\varphi}^2}{aM^6}\nonumber \\&+\bigg[\frac{1}{30} \lambda ^2 (7-60 \xi) \varphi ^2\bigg]\frac{\da^2\ddot{\varphi}}{a^2M^4}+\bigg[\frac{1}{30} \lambda ^2 \varphi  \left(13 \lambda  \varphi ^2-18 m^2\right)\bigg]\frac{\dot{a} \dot{\varphi}\ddot{\varphi}}{a M^6}+\bigg[\frac{1}{5} \lambda ^2 (1-10 \xi) \varphi ^2\bigg]\frac{\ddot{\varphi } \dda}{aM^4}\nonumber \\&+\bigg[\frac{1}{120} \lambda ^3 \varphi  \left(\lambda ^2 \varphi ^4+12 m^4-32 \lambda  m^2 \varphi ^2\right)\bigg]\frac{\dot{ \varphi}^4}{M^{10}}+\bigg[-\frac{1}{15} \lambda ^3 \varphi ^2 \left(\lambda  \varphi ^2-9 m^2\right)\bigg]\frac{\dot{\varphi }^2 \ddot{\varphi }}{M
 ^8}\nonumber\\&+\bigg[-\frac{1}{10} \lambda ^2 (20 \xi -3) \varphi ^2\bigg]\frac{a^{(3)} \dot{\varphi }}{a M^4}+\bigg[-\frac{1}{5} \lambda ^2 \varphi ^2\bigg]\frac{\dot{a} \varphi ^{(3)}}{a M^4}+\bigg[\frac{1}{15} \lambda ^2 \varphi  \left(\lambda  \varphi ^2-2 m^2\right)\bigg]\frac{\dot{\varphi } \varphi ^{(3)}}{M^6}\bigg\}.}
Covariant conservation was explicitly checked. We also see that at order 0 and 2, the derivatives of $a$ nicely organize themselves as powers of $H=\dot{a}/a$ and scalar curvature $R$. In general, it must be possible to write the result in terms of covariant tensors to the extent that the gradient truncation includes all derivatives of a given order.

\subsection{The conformal anomaly}
\label{sec:confa}
As was pointed out at the end of subsection \ref{sec:eomenergy}, in the massless limit in a one-loop calculation we can formally work our way around the infrared singularity at $m=0$ by renormalizing $\lambda$ at a non-zero scale for the field $\varphi$. So in the conformal limit $\xi =\f{n-2}{4(n-1)}$ and $m=0$, we can use the counterterms defined in (\ref{eq:counter1} - \ref{eq:cond2}) with the exception of $\delta\lambda$, which we define as
\ee{\label{eq:conla}\f{\partial^4\langle\hat{T}_{00}\rangle}{\partial\varphi^4}\bigg\vert_{\varphi=\mu,a=1}=\lambda\quad \Leftrightarrow\quad\delta\lambda=\int \f{d^{n-1}k}{2(2\pi)^{n-1}}\bigg\{\f{3\lambda^2(\lambda\mu^2-2\omega_\mu^2)(5\lambda\mu^2-2\omega_\mu^2)}{16\omega_\mu^7}\bigg\},}
where $\omega_\mu=\mathbf{k}^2+m^2+\lambda\mu^2/2$. Inserting these into equation (\ref{eq:trace}) we get the result
\ee{\label{eq:confa}\langle\hat{T}_\mu^{~\mu}\rangle=\frac{\lambda ^2 \varphi ^4}{128 \pi ^2}+\frac{\lambda  \varphi  \ddot{\varphi }}{96 \pi^2}+\frac{\lambda  \dot{\varphi }^2}{96 \pi ^2}+\frac{ \lambda  \dot{\varphi } \varphi\dot{a} }{32 \pi ^2 a}-\frac{a^{(4)}}{480 \pi ^2 a}+\frac{\dot{a}^2 \ddot{a}}{160 \pi ^2 a^3}-\frac{\ddot{a}^2}{480 \pi ^2 a^2}-\frac{a^{(3)} \dot{a}}{160 \pi ^2 a^2},}
which coincides with \cite{BoschiFilho:1991xz}. We point out that the conformal anomaly can also be derived from the renormalized four-dimensional components of $T_{\mu\nu}$. This calculation can be found in appendix \ref{sec:appconf}.

\section{Comparison with other methods}
\label{sec:comparison}
\subsection{Comparison with the derivative expansion}
\label{sec:derivative}
Since we have used an adiabatic expansion for the modes in obtaining the results (\ref{eq:EOMRRR1} - \ref{eq:EOMRRR3}) it is illuminating to compare the results of the previous section to the equations of motion derived via the gradient or Schwinger-DeWitt expansion \cite{Minakshisundaram:1949xg,DeWitt:1965jb}, which is also in essence adiabatic. It is believed that results derived via the Schwinger-DeWitt expansion for the effective action and the adiabatic expansion for the modes in fact give identical results. For a more detailed exposition of this method we refer the reader to \cite{Markkanen:2012rh}. 

We start by writing the standard loop expansion for the effective action
\begin{equation}
\Gamma[\varphi,g^{\mu\nu}]=\int d^n x \sqrt{-g} \mathcal{L}_{eff}=\Gamma^{(0)}[\varphi,g^{\mu\nu}]+\Gamma^{(1)}[\varphi,g^{\mu\nu}]+\cdots,
\end{equation}
where for (\ref{eq:actm}) and (\ref{eq:actg}) we have
\begin{equation}
\label{eq:gammas}
\Gamma^{(0)}[\varphi,g^{\mu\nu}]=S_m[\varphi,g^{\mu\nu}]_0+ S_g[g^{\mu\nu}]_0,\quad\Gamma^{(1)}[\varphi,g^{\mu\nu}]=\f{i}{2}\text{Tr log}~G^{-1}(x,x'),
\end{equation}
and $G(x,x')$ is the Feynman propagator satisfying
\begin{equation}
\bigg[-\Box+m^2+\xi R +\lambda\f{\varphi^2}{2}\bigg]G(x,x')=\f{\delta(x-x')}{\sqrt{-g}}.
\end{equation}
The dimensionally regularized derivative expansion for the one-loop contribution reads
\begin{equation}\label{series expansion}
\Gamma^{(1)}[\varphi,g^{\mu\nu}]=\int d^nx \sqrt{-g}~\f{1}{2(4\pi)^{n/2}}\bigg(\f{M'}{\mu}\bigg)^{n-4}\sum^\infty_{k=0}M'^{4-2k}a_k(x,x)\Gamma(k-n/2),
\end{equation}
where \ee{M'^2=m^2+\bigg(\xi-\f{1}{6}\bigg)R+\lambda\f{\varphi^2}{2},}
and the scale $\mu$ was introduced to maintain proper dimensions for the $n$-dimensional action. We will calculate the result only up to two derivatives (i.e. to second adiabatic order) and for this we need the Schwinger-DeWitt coefficients from $a_0$ to $a_3$ that can be found in appendix \ref{sec:CandSD}. The result is
\begin{align}
\label{aa2}
\Gamma[\varphi,g^{\mu\nu}]&=S_m[\varphi,g^{\mu\nu}]_0+S_g[g^{\mu\nu}]_0\nonumber \\&+\int d^nx \sqrt{-g}~\f{1}{64\pi^2}\bigg\{M'^4\bigg[\f{3}{2}-\log\bigg(\f{M'^2}{\tilde{\mu}^2}\bigg)\bigg]+\f{\lambda\Box\varphi^2}{6}\log\bigg(\f{M'^2}{\tilde{\mu}^2}\bigg)+\f{\lambda^2\nabla^\mu\varphi^2\nabla_\mu\varphi^2}{24M'^2}\bigg\},
\end{align}
where
\ee{\log(\tilde{\mu}^2)=\f{2}{4-n}-\gamma_e+\log(4\pi{\mu}^2).} 
Using the renormalization conditions
\begin{align}
\label{sec:SDrenom}\f{\partial^2\mathcal{L}_{eff}}{\partial\varphi^2}\bigg\vert_{\varphi=0,~g^{\mu\nu}=\eta^{\mu\nu}}&=-m^2, &\f{\partial^4\mathcal{L}_{eff}}{\partial\varphi^4}\bigg\vert_{\varphi=0,~g^{\mu\nu}=\eta^{\mu\nu}}&=-\lambda,\nonumber \\ \mathcal{L}_{eff}\Big\vert_{\varphi=0,~g^{\mu\nu}=\eta^{\mu\nu}}&=\Lambda,&\f{\partial\mathcal{L}_{eff}}{\partial R}\bigg\vert_{\varphi=0,~g^{\mu\nu}=\eta^{\mu\nu}}&=\alpha,
\end{align}
and discarding terms of $\mathcal{O}(R^2)$ we can derive the counterterms, which can be found in appendix \ref{sec:CandSD}. Inserting these into (\ref{aa2}) gives the effective Lagrangian
\eg{\mathcal{L}_{eff}[\varphi,g^{\mu\nu}]=\Lambda+\alpha R-\f{1}{2}\bigg[\partial_\mu\varphi\partial^\mu\varphi+m^2\varphi^2 +\xi R\varphi^2+2\f{\lambda}{4!}\varphi^4\bigg]\nonumber \\+~\f{1}{64\pi^2}\bigg\{\frac{1}{24} \left[9 \lambda ^2 \phi ^4+6 \lambda  \phi ^2 \left(2 m^2+(6 \xi -1) R\right)+4 m^2 (6 \xi -1) R\right]+\f{\lambda^2\varphi^2\nabla_\mu\varphi\nabla^\mu\varphi}{6M'^2}\bigg\}\nonumber \\+~\bigg(-M'^4+\f{\lambda}{6}\Box\varphi^2\bigg)\log\bigg(\f{M'^2}{m^2}\bigg)\bigg\}+\mathcal{O}(A^{(4)}),}
where the last term symbolises the neglected fourth adiabatic order. From this expression the equations of motion can be derived by variation \ee{\f{\delta}{\delta g^{\mu\nu}}\int d^4x\sqrt{-g}~\mathcal{L}_{eff}[\varphi,g^{\mu\nu}]=0,\quad\f{\delta}{\delta \varphi}\int d^4x\sqrt{-g}~\mathcal{L}_{eff}[\varphi,g^{\mu\nu}]=0.} If we choose $g_{\mu\nu}=diag(-1,a^2,a^2,a^2)$ the results are precisely those in (\ref{eq:EOMRR1}), (\ref{eq:EOMRR2}) and (\ref{eq:EOMRR3}) so to this order the Schwinger-DeWitt expansion for the action gives identical results to the adiabatic expansion of modes. It is also plausible that this equality holds as well for higher order results, but of course the proof of this assumption requires higher order calculations.
\subsection{Comparison with adiabatic subtraction}
\label{sec:compsubtraction}
In this subsection we will focus on the time components of the energy-momentum tensor. Let us first write the finite parts of the subtraction term $\delta T_{\mu\nu}$ from (\ref{eq:renT}) by using the counterterms (\ref{eq:counter1} - \ref{eq:cond2})
\ea{\label{eq:comc}(\delta T_{00})_{\rm finite}&=-\f{1}{64\pi^2}\bigg\{{m^4}+{\lambda m^2}\varphi^2+\f{\lambda^2}{4}\varphi^4+2(n-1)\f{\lambda(6\xi-1)}{3}\varphi\dot{\varphi}\f{\da}{a}\nonumber \\&+(n-1)\bigg(\f{n}{2}-1\bigg)\f{\lambda(6\xi-1)}{3}\varphi^2\bigg(\f{\da}{a}\bigg)^2+2{m^2(6\xi-1)}\bigg(\f{\da}{a}\bigg)^2\nonumber \\&+(1-6 \xi )^2 \left(-\frac{\dda^2}{a^2}-3\frac{\da^4}{a^4}+2\frac{ a^{(3)} \da}{a^2}+2\frac{ \da^2 \dda}{a^3}\right)\bigg\}\log\bigg(\f{m^2}{\mu^2}\bigg)+\cdots,}
where the dots indicate contributions that vanish in the four dimensional limit. We have introduced an arbitrary scale $\mu$ for dimensional purposes and used it to absorb a factor of $4\pi$ and the Euler gamma constant from the $n$-dimensional integrals. We explicitly see that in the one-loop approximation the finite part of the subtraction term $\delta T_{00}$ is formed solely from the terms in the classical action (\ref{eq:actm} - \ref{eq:actg}) i.e. it is a polynomial in the degrees of freedom, $\varphi$, $\da/a$ etc., as it is apparent from the derivation in subsection (\ref{sec:energy}).

In a similar fashion we can write explicitly the finite logarithmic parts for the energy-momentum tensor calculated in the fourth order adiabatic vacuum
\ea{\label{eq:comd}-\langle \hat{T}^Q_{00}\rangle_{\rm finite}&=-\f{1}{64\pi^2}\bigg\{{m^4}+{\lambda m^2}\varphi^2+\f{\lambda^2}{4}\varphi^4+2(n-1)\f{\lambda(6\xi-1)}{3}\varphi\dot{\varphi}\f{\da}{a}\nonumber \\&+(n-1)\bigg(\f{n}{2}-1\bigg)\f{\lambda(6\xi-1)}{3}\varphi^2\bigg(\f{\da}{a}\bigg)^2+2{m^2(6\xi-1)}\bigg(\f{\da}{a}\bigg)^2\nonumber \\&+(1-6 \xi )^2 \left(-\frac{\dda^2}{a^2}-3\frac{\da^4}{a^4}+2\frac{ a^{(3)} \da}{a^2}+2\frac{ \da^2 \dda}{a^3}\right)\bigg\}\log\bigg(\f{M^2}{\mu^2}\bigg)+\cdots}
All the finite terms not included in (\ref{eq:comd}) are proportional to a finite inverse power of $M^2$ and vanish at $\lambda=0$. In adiabatic subtraction $-\langle \hat{T}^Q_{00}\rangle$ is precisely the term one uses to render the bare energy-momentum tensor finite as in (\ref{eq:adsub}). Comparing (\ref{eq:comc}) with (\ref{eq:comd}) we see that there is a discrepancy between the two expressions. In particular in (\ref{eq:comd}) there are terms that are not polynomials in $\varphi$ and hence cannot be obtained from the classical action by redefining its constants, which can be seen from the $M^2=m^2+\lambda\varphi^2/2$ dependence of the logarithm. When one sets $\lambda=0$ we have $M\rightarrow m$ and the two expressions coincide verifying \cite{Bunch:1980vc}. Hence we conclude that adiabatic subtraction coincides with the method described in subsection (\ref{sec:eomenergy}) only at the non-interacting limit, taking all renormalization scales to zero for the gravitational counterterms. In fact this result could have been easily deduced from the expressions in (\ref{eq:EOMRRR1} - \ref{eq:EOMRRR3}): We chose the fourth adiabatic order vacuum for the bare energy-momentum tensor, which means that in the context of adiabatic subtraction, the bare expressions are equal to the subtraction terms in (\ref{eq:adsub}) and hence adiabatic subtraction gives identically zero results for all the quantum contributions. This is the case for (\ref{eq:EOMRRR1} - \ref{eq:EOMRRR3}) only when $\lambda=0$.

\section{Summary and Conclusion}
\label{sec:Summary} 
We have shown how to derive the renormalization counterterms consistently from the components of the energy-momentum tensor in curved space-time. This procedure also allows control over the finite parts of the counterterms. When working at the equation of motion level one is free to constrain the metric to be of the desired form. We then used our method for calculating, in the one-loop approximation, the renormalized equation of motion for the field and the components of the energy momentum-tensor to fourth adiabatic order in the adiabatic vacuum. We also calculated the result for the anomalous trace, in the one-loop approximation and finally compared our results to those obtained via two standard methods: The gradient expansion and adiabatic subtraction.
 
We find that the method proposed here has advantages over both the gradient expansion and adiabatic subtraction for the reasons just described: Being able to constrain the metric allows us to specialize and solve for a wider range of models and explicit knowledge of the finite parts of the counterterms is important for the physical interpretation of the coupling constants. The steps described in subsection \ref{sec:eomenergy} are completely general and should be applicable to various different vacuum states 
and different theories. We chose to use our method for the adiabatic vacuum, since this provided fruitful comparisons to the gradient expansion and adiabatic subtraction. We were able to verify that our method gives identical results to the Schwinger-DeWitt expansion to second adiabatic order, which likely indicates an equivalence in higher orders as well.

The comparison with adiabatic subtraction revealed that when interactions are included, adiabatic subtraction introduces counterterms that cannot be obtained from redefining the coupling constants of the theory alone. Our procedure has no such problems. For the non-interacting case this discrepancy is absent.

To what extent a one-loop, fourth order gradient expansion of the effective action is a reliable approximation to describe the dynamics of a quantum field during inflation remains to be quantified. Our choice of example vacuum also allows us for the moment to side-step a long standing issue of how to renormalize to a vacuum with a non-zero cosmological constant (see for instance \cite{Sola:2013gha} for a recent review). It is likely that the renormalization scale must be chosen closer to the inflationary scale, in which case one must perhaps consider a different vacuum to expand around than Minkowski. This is the topic of future work. 

When using this formalism in situations where the energy-momentum tensor involves convolutions over the various degrees of freedom, such as higher order loop expansions, additional complications may arise since the energy-momentum tensor is no-longer a simple function of the degrees of freedom, and a mode function anzats for the field may be inapplicable. In addition, non-analyticities in the renormalization scale choices may pose further obstacles. Nevertheless we believe that this method will be usable in various applications of curved space field theory due to its generality and mathematical straightforwardness.

\section*{Acknowledgements}
We would like to thank Mark Hindmarsh, Keijo Kajantie and Kari Rummukainen for stimulating discussions. This work was supported by the Finnish Academy of Science and Letters and Academy of Finland through project number 1134018.

\appendix
\section{Solutions for the modes up to four derivatives}
\label{sec:Sol}
From (\ref{eq:W}) we find the zeroth order solution of $W^2$ in (\ref{eq:Wexp}) to be 
\ee{(W^2)^{(0)}=\omega^2\label{eq:W0}.}
The higher non-zero contributions are
\ea{(W^2)^{(2)}&=\frac{\dda}{a}\left[\frac{1}{2}\big(2-4\xi+n(4\xi-1)-\g\big)\right]+\left(\frac{\dot{a}}{a}\right)^{2}\bigg[\f{1}{4}\Big((n-2)\big(2-4\xi+n(4\xi-1)\big)\nonumber \\&\phantom{~\frac{\dda}{a}\bigg[\frac{1}{2}}-4\g+5\g^2\Big) \bigg]+\f{\da}{a}\f{\dot{M}}{M}\bigg[\f{5}{2}\gamma^2-\f{5}{2}\g\bigg]+\bigg(\f{\dot{M}}{M}\bigg)^2\bigg[\f{5}{4}\g^2-\f{1}{2}\g\bigg]-\f{\ddot{M}}{M}\f{\g}{2}\label{eq:W2} \nonumber \\&}and
\ea{\label{eq:W4} &(W^2)^{(4)}\nonumber\\&=\f{\da^4}{a^4\omega^2}\bigg\{\bigg[n+n^2\bigg(\xi-\f{1}{4}\bigg)+2\xi-3n\xi\bigg]\g-\f{1}{8}\bigg[8(9+5\xi)\nonumber \\&\phantom{+\f{\da^4}{a^4\omega^2}\bigg\{\bigg]n~}+5n\big(4-n+4(n-3)\xi\big)\bigg]\g^2+27\g^3-\f{135}{8}\g^4\bigg\}\nonumber \\&+\f{\dda\da^2}{a^3\omega^2}\bigg\{-\f{1}{8}\big(n-1)\big(2-n+4(n-1)\xi\big)+\f{3}{8}\bigg[9+4\xi +n\big(5-12\xi+n(8\xi-2)\big)\bigg]\g\nonumber \\&\phantom{+\f{\da^4}{a^4\omega^2}\bigg\{\bigg]n~}+\f{1}{4}\bigg[5n-77-20(n-1)\xi\bigg]\g^2+\f{27}{2}\g^3\bigg\}\nonumber \\&+\f{\da a^{(3)}}{a^2\omega^2}\bigg\{-\f{1}{8}\big(n+1\big)\big(2-n+4(n-1)\xi\big)+\f{1}{8}\bigg[21-5n +20 (n-1)\xi\bigg]\g-\f{7}{4}\g^2\bigg\}\nonumber \\&+\f{\dda^2}{a^2\omega^2}\bigg\{-\f{1}{8}\big(n-1\big)\big(2-n+4(n-1)\xi\big)+\f{1}{8}\bigg[11-2n+8(n-1)\xi\bigg]\g-\f{9}{8}\g^2\nonumber \\&+\f{a^{(4)}}{a\omega^2}\bigg\{\f{1}{8}\bigg[n-2+4\xi(1-n)+\g\bigg]\bigg\}\nonumber \\&+\f{\da^3\dot{M}}{a^3M\omega^2}\bigg\{\f{5}{8}\bigg[9+n(4-12\xi)+8\xi+n^2(-1+4\xi)\bigg]\g\nonumber \\&\phantom{+\f{\da^4}{a^4\omega^2}\bigg\{\bigg]n~}-\f{1}{4}\bigg[231+40\xi+5n\big(4-n+4(-3+n)\xi\big)\bigg]\g^2+\f{243}{2}\g^3-\f{135}{2}\g^4\bigg\}\nonumber \\&+\f{\da^2\dot{M}^2}{a^2M^2\omega^2}\bigg\{\f{1}{8}\bigg[51+n(4-12\xi)+8\xi+n^2(-1+4\xi)\bigg]\g\nonumber \\&\phantom{+\f{\da^4}{a^4\omega^2}\bigg\{\bigg]n~}-\f{1}{8}\bigg[633+n(20-60\xi)+40\xi+5n^2(-1+4\xi)\bigg]\g^2+\f{351}{2}\g^3-\f{405}{4}\g^4\bigg\}\nonumber \\&+\f{\dda\da\dot{M}}{a^2M\omega^2}\bigg\{\f{5}{8}\bigg[18+n-4\xi+n^2(-1+4\xi)\bigg]\g+\f{1}{4}\bigg[-163+10n-40(-1+n)\xi\bigg]\g^2+27\g^3\bigg\}\nonumber \\&+\f{\dda\dot{M}^2}{a^2M^2\omega^2}\bigg\{\f{1}{4}\bigg[10-n+4(-1+n)\xi\bigg]\g+\f{1}{4}\bigg[-68+5n-20(-1+n)\xi\bigg]\g^2+\f{27}{2}\g^3\bigg\}\nonumber \\&+\f{\da^2\ddot{M}}{a^2M\omega^2}\bigg\{\f{1}{8}\bigg[51+8\xi+n\big(4-n+4(-3+n)\xi\big)\bigg]\g-\f{81}{4}\g^2+\f{27}{2}\g^3\bigg\}\nonumber \\&+\f{\da\dot{M}\ddot{M}}{aM^2\omega^2}\bigg\{\f{21}{4}\g-
 \f{129}{4}\g^2+27\g^3\bigg\}+\f{\dda\ddot{M}}{aM\omega^2}\bigg\{\f{1}{4}\bigg[10-n+4(-1+n)\xi\bigg]\g-\f{9}{4}\g^2\bigg\}\nonumber \\&+\f{\da\dot{M}^3}{aM^3\omega^2}\bigg\{-27\g^2+\f{189}{2}\g^3-\f{135}{2}\g^4\bigg\}+\f{a^{(3)}\dot{M}}{aM\omega^2}\bigg\{\f{1}{8}\bigg[19-5n+20(n-1)\xi\bigg]\g-\f{7}{4}\g^2\bigg\}\nonumber \\&+\f{\da M^{(3)}}{aM\omega^2}\bigg\{\f{7}{4}\g-\f{7}{4}\g^2\bigg\}+\f{\dot{M}^{4}}{M^4\omega^2}\bigg\{-\f{9}{8}\g^2+\f{27}{2}\g^3-\f{135}{8}\g^4\bigg\}+\f{M^{(4)}}{M\omega^2}\bigg\{\f{1}{8}\g\bigg\}\nonumber \\&+\f{\dot{M}^2\ddot{M}}{M^3\omega^2}\bigg\{-\f{15}{2}\g^2+\f{27}{2}\g^3\bigg\}+\f{\ddot{M}^2}{M^2\omega^2}\bigg\{\f{3}{8}\g-\f{9}{8}\g^2\bigg\}+\f{\dot{M}{M}^{(3)}}{M^2\omega^2}\bigg\{\f{1}{2}\g-\f{7}{4}\g^2\bigg\},}
where $M^2=m^2+(\lambda/2)\varphi^2$, $\omega^2=\mathbf{k}^2/a^2+M^2$ and $\gamma = M^2/\omega^2$.
\section{Counterterms for the adiabatic energy-momentum tensor}
\label{sec:count}
Using the renormalization conditions from (\ref{cond}) for the quantum energy-momentum tensor components (\ref{eq:T00}) and (\ref{eq:Tii}) calculated with the fourth order adiabatic modes, we get the counterterms
\ea{\label{eq:counter1}\delta m^2&=\int \f{d^{n-1}k}{2(2\pi)^{n-1}}\bigg\{-\f{\lambda}{2\omega_0}\bigg\}\\\delta \lambda&=\int \f{d^{n-1}k}{2(2\pi)^{n-1}}\bigg\{\f{3\lambda^2}{4\omega_0^3}\bigg\}\\\delta \xi&=\int \f{d^{n-1}k}{2(2\pi)^{n-1}}\bigg\{-\f{\lambda\Big[m^2+\big(-2+n+4\xi(1-n)\big)\omega_0^2\Big]}{16(n-1)\omega_0^5}\bigg\} \\\delta \Lambda&=\int \f{d^{n-1}k}{2(2\pi)^{n-1}}\bigg\{\omega_0\bigg\}\\\delta \alpha&=\int \f{d^{n-1}k}{2(2\pi)^{n-1}}\bigg\{\f{m^4-\omega_0^2\big[2-n+4(n-1)\xi\big]\big[2m^2+(n-2)\omega_0^2\big]}{8\big[2+(n-3)n\big]\omega_0^5}\bigg\}\label{eq:counter5},\\\delta\beta&=\f{n\delta\epsilon_1+4\delta\epsilon_2} {4(1-n)}+\int \f{d^{n-1}k}{2(2\pi)^{n-1}}\bigg\{-\f{\big(m^2+(n-2+4\xi(1-n))\omega_0^2\big)^2}{128(n-1)^2\omega_0^7}\bigg\} \\  \delta\epsilon_1&=\f{4\delta\epsilon_2}{2-n}+\int \f{d^{n-1}k}{2(2\pi)^{n-1}}\bigg\{\frac{1}{32(n-4)(n-2)^2(n-1)\omega_0^{11}}\bigg[\nonumber \\&+105 m^8-28 m^6 (20 (n-1) \xi -5 n+14) \omega_0^2\nonumber \\&+m^4 \left(9 n^2-40 (n-18) (n-1) \xi -188 n+356\right) \omega_0^4\nonumber \\&-2 m^2 (4 (n-1) \xi -n+2) (24 (n-3)
   \xi  n-7 (n-4) n+48 \xi -12) \omega_0^6\nonumber \\&-4 (n-4) (n-2) (-4 \xi  n+n+4 \xi -2)^2 \omega_0^8\bigg]\bigg\}\label{eq:cond2}.}
where $\omega_0^2=\mathbf{k}^2+m^2$. The surprising thing is that $\delta \epsilon_2$ is not needed: when $\delta \beta$ and $\delta \epsilon_2$ are inserted into $\delta T^{g}_{\mu\nu}$ the coefficient of $\delta \epsilon_2$ vanishes identically.

\section{Geometric tensors in $n$ dimensional FRW spaces}
\label{sec:appA}
Standard variational calculus gives the following geometric tensors
\begin{align}
G_{\mu\nu}\equiv\f{1}{\sqrt{-g}}\f{\delta}{\delta g^{\mu\nu}}\int d^nx\sqrt{-g}~R=-\f{1}{2} Rg_{\mu\nu}+R_{\mu\nu},
\end{align}
\begin{align}
\f{1}{\sqrt{-g}}\f{\delta}{\delta g^{\mu\nu}}\int d^nx\sqrt{-g}~Rf(x)=\big[-\f{1}{2} Rg_{\mu\nu}+R_{\mu\nu}-\nabla_\mu\nabla_\nu+g_{\mu\nu}\Box\big]f(x),
\end{align}
\begin{align}
~^{(1)}H_{\mu\nu}\equiv\f{1}{\sqrt{-g}}\f{\delta}{\delta g^{\mu\nu}}\int d^nx\sqrt{-g}~R^2=-\f{1}{2} R^2g_{\mu\nu}+2R_{\mu\nu}R-2\nabla_\mu\nabla_\nu R+2g_{\mu\nu}\Box R,
\end{align}
\begin{align}
~^{(2)}H_{\mu\nu}&\equiv\f{1}{\sqrt{-g}}\f{\delta}{\delta g^{\mu\nu}}\int d^nx\sqrt{-g}~R^{\mu\nu}R_{\mu\nu}\nonumber \\&=-\f{1}{2}R_{\alpha\beta}R^{\alpha\beta}g_{\mu\nu}+ 2R_{\rho\nu\gamma\mu}R^{\rho\gamma}-\nabla_\nu\nabla_\mu R+\f{1}{2}\Box R g_{\mu\nu}+\Box R_{\mu\nu},
\end{align}
and
\begin{align}
H_{\mu\nu}&\equiv\f{1}{\sqrt{-g}}\f{\delta}{\delta g^{\mu\nu}}\int d^nx\sqrt{-g}~R^{\mu\nu\sigma\delta}R_{\mu\nu\sigma\delta}\nonumber \\&=
-\f{g_{\mu\nu}}{2}R^{\alpha\sigma\gamma\delta}R_{\alpha\sigma\gamma\delta} +2{R_\mu}^{\rho\alpha\sigma}R_{\nu\rho\alpha\sigma}+4R_{\sigma\mu\gamma\nu}R^{\gamma\sigma}- 4R_{\mu\gamma}{R^\gamma}_{\nu}+4\Box R_{\mu\nu}-2\nabla_\mu\nabla_\nu R.
\end{align}
The $n$ dimensional formulae can be derived by noting that irrespective of the dimensionality, the non-zero components of the Riemann tensor in conformal time,\newline $g_{\mu\nu}dx^\mu dx^\nu=a(\eta)(-d\eta^2+d\mathbf{x}^2)$, are
\ee{{R^\eta}_{i\eta i}=\f{a''}{a}-\bigg(\f{a'}{a}\bigg)^2=-{R^i}_{\eta i\eta }\quad\text{and}\quad{R^j}_{ij i}=\bigg(\f{a'}{a}\bigg)^2,}
and hence we have
\ea{\label{eq:R}R_{\eta\eta}&=(n-1)\bigg[\bigg(\f{a'}{a}\bigg)^2-\f{a''}{a}\bigg],\quad R_{ii}=\f{a''}{a}+(n-3)\bigg(\f{a'}{a}\bigg)^2,\nonumber \\R&=\f{1}{a^2}\bigg[2(n-1)\f{a''}{a}+(n-1)(n-4)\bigg(\f{a'}{a}\bigg)^2\bigg].
}
Using the above expressions we can now write our tensors in Minkowski time. The second order tensors
\ea{(-\nabla_0\nabla_0+g_{00}\Box)f(t)&=(n-1)\f{\da}{a}\dot{f}(t), \\
(-\nabla_i\nabla_i+g_{ii}\Box)f(t)&=a^2\bigg[(2-n)\f{\da}{a}\dot{f}(t)-\ddot{f}(t)\bigg],\\
R&=2(n-1)\bigg(\f{\da^2}{a^2}+\f{\dda}{a}\bigg)+(n-1)(n-4)\f{\da^2}{a^2}\label{eq:Rt},\\
G_{00}&=\f{(n-1)(n-2)}{2}\bigg(\f{\dot{a}}{a}\bigg)^2,\\
G_{ii}&=a^2(2-n)\bigg[\f{(n-3)}{2}\bigg(\f{\dot{a}}{a}\bigg)^2+\f{\ddot{a}}{a}\bigg].}
and the fourth order tensors
\ea{ ~^{(1)}H_{00}&=\f{1}{2}(n-10)(n-2)(n-1)^2\f{\dot{a}^4}{a^4}+4(n-3)(n-1)^2\f{\dot{a}^2\ddot{a}}{a^3}-2(n-1)^2\f{\ddot{a}^2}{a^2}\nonumber\\&+4(n-1)^2\f{\dot{a}a^{(3)}}{a^2},\\
 ~^{(1)}H_{ii}&=a^2\bigg[-\f{1}{2}(n-10)(n-5)(n-2)(n-1)\f{\dot{a}^4}{a^4}-2(n-1)\big(44+n(3n-26)\big)\f{\dot{a}^2\ddot{a}}{a^3}\nonumber \\&-6\big(3+(n-4)n\big)\f{\ddot{a}^2}{a^2}-8\big(3+(n-4)n\big)\f{\dot{a}a^{(3)}}{a^2}+4(1-n)\f{a^{(4)}}{a}\bigg],\\ 
 ~^{(2)}H_{00}&=-\f{1}{2}(n^2-4)(n-1)\f{\dot{a}^4}{a^4}+n\big(3+(n-4)n\big)\f{\dot{a}^2\ddot{a}}{a^3}-(n-1)\f{n}{2}\f{\ddot{a}^2}{a^2}+(n-1)n\f{\dot{a}a^{(3)}}{a^2},\\ 
 ~^{(2)}H_{ii}&=a^2\bigg[\f{1}{2}(n-5)(n^2-4)\f{\dot{a}^4}{a^4}-\Big[8+n\big(12+(n-9)n\big)\Big]\f{\dot{a}^2\ddot{a}}{a^3}\nonumber \\&-\f{3(n-3)n}{2}\f{\ddot{a}^2}{a^2}-2(n-3)n\f{\dot{a}a^{(3)}}{a^2}-n\f{a^{(4)}}{a}\bigg],\\ 
 H_{00}&=-3\big(2+(n-3)n\big)\f{\dot{a}^4}{a^4}+4\big(3+(n-4)n\big)\f{\dot{a}^2\ddot{a}}{a^3}+2(1-n)\f{\ddot{a}^2}{a^2}+4(n-1)\f{\dot{a}a^{(3)}}{a^2},\\ 
 H_{ii}&=a^2\bigg[3(n-5)(n-2)\f{\dot{a}^4}{a^4}-4\big(18+(n-10)n\big)\f{\dot{a}^2\ddot{a}}{a^3}-6(n-3)\f{\ddot{a}^2}{a^2}-8(n-3)\f{\dot{a}a^{(3)}}{a^2}\nonumber\\&-4\f{a^{(4)}}{a}\bigg].
 }
\section{Conformal anomaly from the renormalized $T_{\mu\nu}$}
\label{sec:appconf}
If we repeat the calculation of section \ref{sec:expli}, but with renormalizing the $\lambda$ at a non-zero scale as in (\ref{eq:conla}) and set $\xi=1/6$ and $m=0$ we get
\ea{\langle\hat{T}_{00}\rangle&=\f{1}{2}\dot{\varphi}^2+\f{\lambda}{4!}\varphi^4+\f{\varphi^2}{2}\f{\da^2}{a^2}+\varphi\dot{\varphi}\f{\da}{a}+\f{1}{64\pi^2}\bigg\{\f{\lambda^2\varphi^4}{4}\bigg[\log\bigg(\f{\varphi^2}{\mu^2}\bigg)-\f{25}{6}\bigg]+\f{\lambda}{3}\dot{\varphi}^2\nonumber \\&+\f{2}{5}\f{\da^3\dot{\varphi}}{a^3\varphi}+\f{17}{15}\f{\da^2\dot{\varphi}^2}{a^2\varphi^2} +\f{8}{15}\f{\da\dot{\varphi}^3}{a\varphi^3}-\f{1}{15}\f{\dot{\varphi}^4}{\varphi^4}-\f{2}{15}\f{\da\dda\dot{\varphi}}{a^2\varphi}-\f{4}{15}\f{\dda\dot{\varphi}^2}{a\varphi^2}-\f{2}{5}\f{\da^2\ddot{\varphi}}{a^2\varphi}\nonumber \\&+\f{2}{15}\f{\da\dot{\varphi}\ddot{\varphi}}{a\varphi^2} +\f{4}{15}\f{\dot{\varphi}^2\ddot{\varphi}}{\varphi^3}+\f{2}{15}\f{\dda\ddot{\varphi}}{a\varphi} +\f{1}{15}\f{\ddot{\varphi}^2}{\varphi^2}-\f{2}{15}\f{\dot{\varphi}a^{(3)}}{a\varphi}-\f{2}{15}\f{\da\varphi^{(3)}}{a\varphi}-\f{2}{15}\f{\dot{\varphi}\varphi^{(3)}}{\varphi^2}\bigg\},}

\ea{a^{-2}\langle\hat{T}_{ii}\rangle&=\f{1}{6}\dot{\varphi}^2-\f{\lambda}{4!}\varphi^4-\f{\varphi^2}{6}\bigg(\f{\da^2}{a^2}+2\f{\dda}{a}\bigg)-\f{2}{3}\varphi\dot{\varphi}\f{\da}{a}-\f{1}{3}\varphi\ddot{\varphi}+\f{1}{64\pi^2}\bigg\{\nonumber \\&-\f{\lambda^2\varphi^4}{4}\bigg[\log\bigg(\f{\varphi^2}{\mu^2}\bigg)-\f{25}{6}\bigg]+\f{\lambda}{3}\dot{\varphi}^2-\f{1}{9}\f{\da^2\dot{\varphi}^2}{a^2\varphi^2}+\f{8}{9}\f{\da\dot{\varphi}^3}{a\varphi^3}
-\f{1}{9}\f{\dot{\varphi}^4}{\varphi^4}-\f{2}{9}\f{\da\dda\dot{\varphi}}{a^2\varphi}\nonumber \\&-\f{4}{9}\f{\dda\dot{\varphi}^2}{a\varphi^2}-\f{10}{9}\f{\da\dot{\varphi}\ddot{\varphi}}{a\varphi^2}+\f{4}{9}\f{\dot{\varphi}\ddot{\varphi}}{\varphi^3}+\f{2}{9}\f{\dda\ddot{\varphi}}{a\varphi}-\f{1}{9}\f{\ddot{\varphi}^2}{\varphi^2}+\f{2}{9}\f{\da\varphi^{(3)}}{a\varphi}-\f{2}{9}\f{\dot{\varphi}\varphi^{(3)}}{\varphi^2}+\f{2}{45}\f{\varphi^{(4)}}{\varphi}\bigg\},
}
and
\ea{\label{eq:eom3c}&\ddot{\varphi}+3\dot{\varphi}\f{\da}{a}+\f{\lambda}{6}\varphi^3+\varphi\bigg(\f{\da^2}{a^2}+\f{\dda}{a}\bigg)+\f{1}{64\pi^2}\bigg\{{\lambda^2\varphi^3}\log\bigg(\f{\varphi^2}{\mu^2}\bigg)-\f{11}{3}\lambda^2\varphi^3+2{\lambda}\dot{\varphi}\f{\da}{a}+\f{2}{3}\lambda\ddot{\varphi}\nonumber \\&-\f{2}{15}\f{\dda^2}{a^2\varphi}+\f{2}{5}\f{\da^3\dot{\varphi}}{a^3\varphi^2}+\f{22}{15}\f{\da^2\dot{\varphi}^2}{a^2\varphi^3} -\f{32}{15}\f{\da\dot{\varphi}^3}{a\varphi^4}+\f{4}{15}\f{\dot{\varphi}^4}{\varphi^5}+\f{8}{15}\f{\da\dda\dot{\varphi}}{a^2\varphi^2}+\f{2}{5}\f{\da^2\dda}{a^3\varphi}+\f{16}{15}\f{\dot{\varphi}^2\dda}{a\varphi^3}-\f{2}{5}\f{\da^2\ddot{\varphi}}{a^2\varphi^2} \nonumber \\&+\f{52}{15}\f{\da\dot{\varphi}\ddot{\varphi}}{a\varphi^3}-\f{16}{15}\f{\dot{\varphi}^2\ddot{\varphi}}{\varphi^4} -\f{8}{15}\f{\dda\ddot{\varphi}}{a\varphi^2}+\f{6}{15}\f{\ddot{\varphi}^2}{\varphi^3}-\f{2}{15}\f{\dot{\varphi}a^{(3)}}{a\varphi^2}-\f{2}{5}\f{\da a^{(3)}}{a^2\varphi}-\f{12}{15}\f{\da \varphi^{(3)}}{a\varphi^2}+\f{8}{15}\f{\dot{\varphi} \varphi^{(3)}}{\varphi^3}\nonumber \\&-\f{2}{15}\f{{\varphi}^{(4)}}{\varphi^2}-\f{2}{15}\f{a^{(4)}}{a\varphi}\bigg\}=0
.}
Denoting the left hand side of (\ref{eq:eom3c}) as $\mathcal{E}$, it can now be used to rewrite the trace of $T_{\mu\nu}$ in various equivalent forms. If we choose the following expression for the trace
\ee{\langle \hat{T}_\mu^{~\mu}\rangle=-\langle \hat{T}_{00}\rangle+\f{3}{a^2}\langle \hat{T}_{ii}\rangle+\varphi\mathcal{E},}
we get exactly the from in (\ref{eq:confa}).
\section{Counterterms and Schwinger-DeWitt coefficients for the effective action} 
\label{sec:CandSD}
The first few $a_0$ terms in the gradient expansion (\ref{series expansion}) can be found in e.g. \cite{Jack:1985mw}
\ea{a_0=1,\quad a_1=0,\quad a_2=-\f{\lambda}{12}\Box\varphi^2+\cdots,\quad a_3=\f{\lambda^2\nabla_\mu\varphi^2\nabla^\mu\varphi^2}{48M'^2}+\cdots,}
where the dots indicate terms with an adiabatic order higher than 2.

From the renormalization conditions in (\ref{sec:SDrenom}) we get the following results:
\begin{align}
\delta m^2&=\f{1}{64\pi^2}\bigg\{2m^2\lambda\bigg[1-\log\bigg(\f{m^2}{\tilde{\mu}^2}\bigg)\bigg]\bigg\},&\delta \lambda &=\f{1}{64\pi^2}\bigg\{-6\lambda^2\log\bigg(\f{m^2}{\tilde{\mu}^2}\bigg)\bigg\}\nonumber, \\ \delta \xi &=\f{1}{64\pi^2}\bigg\{2\lambda\bigg(\f{1}{6}-\xi\bigg)\log\bigg(\f{m^2}{\tilde{\mu}^2}\bigg)\bigg\},&\delta\Lambda &=\f{1}{64\pi^2}\bigg\{\f{m^4_\phi}{2}\bigg[-3+2\log\bigg(\f{m^2}{\tilde{\mu}^2}\bigg)\bigg]\bigg\}\nonumber, \\\delta\alpha &=\f{1}{64\pi^2}\bigg\{2m^2\bigg(\f{1}{6}-\xi\bigg)\bigg[1-\log\bigg(\f{m^2}{\tilde{\mu}^2}\bigg)\bigg]\bigg\},
\end{align}

\end{document}